
\input harvmac.tex
\input epsf.tex

\tolerance = 10000
\noblackbox

%
%
\ifx\epsfbox\UnDeFiNeD\message{(NO epsf.tex, FIGURES WILL BE IGNORED)}
\def\figin#1{\vskip2in}
\else\message{(FIGURES WILL BE INCLUDED)}\def\figin#1{#1}\fi
\def\tfig#1{{\xdef#1{Fig.\thinspace\the\secno-\the\figno}}
Fig.\thinspace\the\secno-\the\figno \global\advance\figno by1}
\def\ifig#1#2#3{\goodbreak\midinsert\figin{\centerline{
{\epsfbox {#3.eps} }}}%
\smallskip\centerline{\vbox{\baselineskip12pt
\hfil\footnotefont{\bf #1:} #2 \hfil}}
\bigskip\endinsert}
\def\gn{\vert G \vert}

\def\C{{\bf C}}
\def\d{\delta}
\def\D{\Delta}
\def\e{\epsilon}
\def\eps{\epsilon}
\def\ie{{\it i.e.}}
\def\eg{{\it e.g.}}
\def\O{{\cal O}}
\def\o#1{{\cal O}_{#1}}
\def\lowdelta#1#2{\Delta_{#2}^{{\phantom{#2}}#1} }
\def\seg#1{${\overline{#1}}$}


\def\vraised{ {\raise1pt\hbox{\font\bigtenrm=cmr10 scaled\magstep1 $v$}} }

\def\sqr#1#2{{\vcenter{\vbox{\hrule height.#2pt
        \hbox{\vrule width.#2pt height#1pt \kern#1pt
                \vrule width.#2pt}
        \hrule height.#2pt \kern1pt}}}}
\def\square{{\mathchoice\sqr{6}4\sqr{6}4\sqr{4}3\sqr{4}3}}

\Title{\vbox{\baselineskip12pt\hbox{\vtop{
          \hbox{CLNS 93/1200} \hbox{ hep-th/9305080 } }}   }}
{{\vbox {\centerline{ Structure of Topological Lattice Field Theories}
\vskip10pt\centerline{ in Three Dimensions}         }}}

\centerline{
{\it Stephen-wei Chung{,}}\footnote{${}^{\natural}$}{
\hbox{\vtop{ \hbox{e-mail:~chung@beauty.tn.cornell.edu\hfil}
          \hbox{Address after Sep. 1, 1993: Fermi National Accelerator
Laboratory,\hfil}
\hbox{P.O. Box 500, Batavia, IL 60510} }}\hfil }
{\it Masafumi Fukuma,}\footnote{${}^{\sharp}$}{
\hbox{\vtop{ \hbox{e-mail:~fukuma@strange.tn.cornell.edu\hfil}
          \hbox{Address after July 1, 1993: Yukawa Institute for
Theoretical Physics,\hfil}
\hbox{Kyoto University, Kitashirakawa, Kyoto 606, Japan} }}\hfil }
{\it and Alfred Shapere}\footnote{${}^{\flat}$}{e-mail:~
         shapere@strange.tn.cornell.edu}  }
\smallskip
\centerline{Newman Laboratory of Nuclear Studies}
\centerline{Cornell University}
\centerline{Ithaca, N.Y.  14853-5001, USA}

\vskip.3in
\medskip
\noindent
We construct and classify topological lattice field theories in three
dimensions. After defining a general class of local lattice field
theories, we impose invariance under arbitrary topology-preserving
 deformations of
the underlying lattice, which are generated by two new local lattice
moves.  Invariant solutions are in one--to--one correspondence with
Hopf algebras satisfying a certain constraint.  As an example, we
study in detail the topological lattice field theory corresponding to
the Hopf algebra based on the group ring $\C[G]$, and show that it is
equivalent to lattice gauge theory at zero coupling, and to the
Ponzano--Regge theory for $G=$SU(2).

\Date{May 1993}

\def\npb#1#2#3{{\it Nucl.\ Phys.} {\bf B#1} (19#2) #3}

\def\mpl#1#2#3{{\it Mod.\ Phys.\ Lett.} {\bf A#1} (19#2) #3}

\def\pr0#1#2#3{{\it Phys.\ Rev.} {\bf #1} (19#2) #3}

\def\cmp#1#2#3{{\it Commun.\  Math.\ Phys.} {\bf #1} (19#2) #3}

\def\jetp#1#2#3{ {\it Sov.\ Phys.\ J.E.T.P.} {\bf #1} (19#2) #3}
\lref\FHK{M.~Fukuma, S. Hosono and H.~Kawai: Lattice
  topological field theory in two-dimensions, Cornell preprint CLNS
  92/1173, hep-th/9212154.}
\lref\Boulatov{D.~Boulatov: \mpl{7}{92}{1629}. }
\lref\TV{V.G.~Turaev and O.Y.~Viro: {\it Topology} {\bf 31} (1992) 865.}
\lref\TURAEV{V.G.~Turaev: Quantum invariants of 3--manifolds,
  Strasbourg preprint ISSN 0755--3390.}
\lref\ALEX{J.W. Alexander: {\it Ann. Math.} {\bf 31} (1930) 292.}
\lref\Abe{E.~Abe: {\it Hopf algebras.} Cambridge: Cambridge University
  Press 1980.}
\lref\Wittenfour{E. Witten: \cmp{141}{91}{435};
  {\it J. Geometry and Phys.} {\bf 9} (1992) 303.}
\lref\Migdal{A. Migdal: \jetp{42}{75}{413}. }
\lref\RP{G. Ponzano and T. Regge: in Bloch, F. (ed.)
  {\it Spectroscopic and Group Theoretical Methods in Physics}.
   Amsterdam: North-Holland  1968.}
\lref\OS{H.~Ooguri and N.~Sasakura: \mpl{6}{91}{3591}; H.~Ooguri,
  \npb{382}{92}{276}.  }
\lref\FG{G.~Felder and O.~Grandjean:
  On combinatorial three-manifold invariants, ETH preprint, 1992.}
\lref\Mat{M. Gross and S. Varsted: \npb{378}{92}{367}.}
\lref\TRIPLE{E. Neher:  {\it Jordan triple systems by the grid approach,
  Lecture Notes in Mathematics.} Berlin: Springer-Verlag, 1987.}
\lref\DONALDSON{E. Witten: \cmp{117}{88}{353}.}
\lref\FOURD{H. Ooguri:  {\it Mod.\ Phys.\ Lett.} {\bf A7} (1992) 2799.}
\lref\ATI{M. Atiyah:  {\it Publ.\ Math.\ IHES} {\bf 68} (1988) 175.}
\lref\CREUTZ{M.\ Creutz: {\it Quarks, Gluons and Lattice.} Cambridge:
 Cambridge University Press, 1983.}
\lref\DJN{B. Durhuus, H.P. Jakobsen and R. Nest:
  {\it Nucl. Phys.} {\bf B\ 25A} {\it (Proc. Suppl.)} (1992) 109.}
\lref\MESS{A. Messiah: {\it Quantum Mechanics,} Vol.\ {\bf II}.
  Amsterdam: North-Holland 1962.}
\lref\BP{C. Bachas and P.M.S. Petropoulos: \cmp{152}{93}{191}.   }
\lref\CSW{E. Witten: {\it Commun.\ Math.\ Phys.} {\bf 121} (1989) 351.}
\lref\Wittenb{E. Witten: {\it Nucl. Phys.} {\bf B340} (1990) 281.}
\lref\Wittenc{E. Witten: {\it Nucl.\ Phys.} {\bf B311} (1988) 46.}
\lref\EY{T. Eguchi and S.K. Yang: {\it Mod.\ Phys.\ Lett.} {\bf A5}
  (1991) 1693.}
\lref\Kontsevich{M. Kontsevich: {\it Commun.\ Math.\ Phys.} {\bf 147}
     (1993) 1.}
\lref\Three{S.~Mizoguchi and T.~Tada: {\it Phys.\ Rev.\ Lett.}
 {\bf 68} (1992) 1795;  F.~Archer and R.M.~Williams:
  {\it Phys.\ Lett.} {\bf B273} (1991) 438.}
\lref\ms{G. Moore and N. Seiberg: {\it Phys.\ Lett.} {\bf B220} (1989) 422.}
\lref\wittend{E. Witten: {\it Nucl.\ Phys.} {\bf B322} (1989) 629;
  E. Witten, {\it Nucl.\ Phys.} {\bf B330} (1990) 285.}
\lref\wittene{E. Witten: {\it Nucl.\  Phys.} {\bf B268} (1986) 253;
  E. Witten, IAS preprint IASSNS--HEP--92--45, hep-th/9207094.}
\lref\JW{J.F.~Wheater: {\it Phys.\ Lett.} {\bf 223B} (1989) 451.}
\lref\TJ{T.~J{\`o}nsson: {\it Phys.\  Lett.} {\bf 265B} (1991) 141.}
\lref\DW{R.~Dijkgraaf and E.~Witten: {\it Commun.\  Math.\  Phys.}
  {\bf 129} (1990) 393.}
\lref\KM{Al.R.~Kavalov and R.L.~Mkatchyan: {\it Phys.\  Lett.}
  {\bf 242B} (1990) 429.}
\lref\FJA{F.J.~Archer: {\it Phys.\  Lett.} {\bf 295B} (1992) 199.}

\newsec{Introduction}
Topological field theories   have played an important
role in our attempts to understand the nonperturbative structure of
string theory and quantum gravity.  In three dimensions, for example,
Witten's observation \Wittenc\ that Einstein gravity can be recast
as a topological field theory has provided us with a solvable model
for testing hypotheses about more realistic theories of gravity.
Three--dimensional Chern--Simons--Witten (CSW) theories have been
useful in understanding and systematizing two--dimensional conformal
field theories \ms\ and integrable models \wittend , and even appear in
formulations of open string field theory \wittene .

Mathematicians have also found topological field theories (TFTs)
 to be valuable tools, particularly
in the study of topological invariants of manifolds.  For example, CSW
theories, in addition to their physical applications,
generate Jones polynomials and other
invariants of three--manifolds as correlation functions \CSW .
Similarly, Donaldson polynomials are generated by a four--dimensional
topological gauge theory \DONALDSON , and topological gravity in two
dimensions models intersection theory on the moduli space of Riemann
surfaces \refs{\Wittenb,\Kontsevich}.

As the list of topological field theories and their applications
continues to grow, so does the need for an axiomatic, constructive
description of them.  In fact, TFTs are so highly constrained by their
symmetry that it does not seem unrealistic to hope that such a
description could lead to a classification of TFTs, or at
least to the discovery of new ones.

The usual description of topological field theories, in terms of
continuum path integrals, possesses the merits of manifest symmetry
and a straightforward semiclassical limit. However, path integral
field theories contain an infinite number of degrees of freedom,
almost all of which must be gauge--fixed or cancelled in order to
obtain a TFT.  Another difficulty of the path integral approach is its
lack of mathematical rigor. Perhaps a simpler formulation, without
spurious degrees of freedom and without divergences, would facilitate
the systematic construction and classification of topological field
theories.

The lattice approach to defining quantum field theories rigorously is
well--established in many contexts.  Usually, the lattice definition
of a quantum field theory requires a continuum limit to be taken, and
to find the appropriate continuum limit is one of the major tasks in
the lattice construction.  However, if the lattice theory is
topological, then all lattice scales are equivalent --- provided the
lattice is fine enough to contain all the topological information
about the underlying manifold --- and in particular, the lattice
theory is formally {\it equivalent} to its continuum limit.
By putting TFTs on the lattice,
calculations of physical quantities are reduced to
combinatorics, no spurious states appear, and, as we aim to show in
this paper, the underlying geometric and algebraic structures
are greatly clarified.

\bigskip

Our general method for constructing TFTs on the lattice will be as
follows.  First we recall that TFTs are continuum field theories
invariant under arbitrary smooth, {\it local\/} deformations of the
background metric.  In such theories, physical quantities depend only
on the topology of the underlying manifold, and not on its local
geometry.  Likewise, a lattice field theory is said to be topological
if it is invariant under arbitrary local deformations of the
underlying lattice.  Thus, in a ``topological lattice field theory''
(TLFT), physical quantities will depend on the lattice only through
its topology. To find a useful criterion for topological invariance,
we first look for a simple set of lattice moves that generate all
local topology--preserving deformations.  Then, to prove that a given
lattice theory is topological, it suffices to show that its partition
function is invariant under these generating moves.

Two--dimensional TLFTs were constructed and extensively studied in
refs.\ \refs{\JW,\TJ,\BP,\FHK} according to the above
procedure.  As will be reviewed in section 2, only two local moves are
needed to generate all topologically equivalent simplicially
triangulated lattices in two dimensions: the bond flip and the bubble
move\ \refs{\FHK,\Mat,\Boulatov}.  Lattice theories invariant under these
moves, \ie\ 2D TLFTs,
are in one--to--one correspondence with semisimple associative
algebras.  The construction is quite general; in fact, all known
topological matter theories in two dimensions ({\it e.g.\/} 2D
topological Yang-Mills theory \Migdal\ and twisted $N=2$ minimal
topological matter \EY) are obtained from TLFTs of this sort
\refs{\BP,\FHK}.

The first three--dimensional TLFT was constructed by Ponzano and Regge
in 1967 \RP, motivated by the so--called Regge calculus of 3D quantum
gravity.  In their model, a tetrahedral lattice is chosen, and each
edge of the lattice is assigned an irreducible representation of
SU(2), so that a 6$j$--symbol of SU(2) can be uniquely determined for
each tetrahedral cell.  The partition function is the sum over
irreducible representations of the product over all tetrahedra of
these 6$j$--symbols, and is shown to be invariant under a set of local
moves that generate all local deformations of the tetrahedral lattice,
namely, barycentric subdivision and a move that transforms two
adjacent tetrahedra into three.  Turaev and Viro \TV\ (see
also\ \refs{\Three,\DJN,\FJA,\FG}) generalized
to models involving $6j$-symbols for the quantum group SU$_q(2)$, and
Dijkgraaf and Witten \DW\ constructed models based on discrete groups.
Subsequently, Turaev \TURAEV, and Ooguri and Sasakura
\OS\ proved that the Ponzano--Regge model is equivalent to
the CSW theory with gauge group ISO$(3)$.
These constructions of 3D TLFTs are strongly based on the properties
of  group (or quantum group) representations, and
moreover, they are only applicable to tetrahedral lattices.  Thus,
they do not seem to encompass all consistent 3D TFTs.

Our main goal in this paper is to give an alternative (and
hopefully more general) approach to 3D TLFTs, with emphasis on the
parallels of our construction with the construction of 2D TLFTs of
refs.\ \refs{\BP,\FHK}.  In particular, we will show that our 3D TLFTs
are in one--to--one correspondence with a class of Hopf algebras.

\bigskip

The present paper is organized as follows.

In section 2, we will review the basic structure of 2D TLFTs
\refs{\BP,\FHK} from a point of view that will carry over in
subsequent sections to higher--dimensional TLFTs.

In section 3, we will present a systematic construction of 3D TLFTs,
defined not only on tetrahedral lattices, but on arbitrary 3D cell
complexes as well.  As in two dimensions, topological invariance is
implemented by a reduced set of two local lattice moves, which are
sufficient to generate all topologically equivalent lattices.  If we
were interested only in tetrahedral lattices, then the two tetrahedral
moves mentioned above would be sufficient. However, in order to account
for arbitrary topologically equivalent lattices, we will introduce a
more powerful set of moves, to be called the ``hinge'' and ``cone''
moves, which actually generate all local deformations of the lattice,
including the original tetrahedral moves.  (The bulk of the proof of
this statement is contained in Appendix B.)  Our construction
of 3D TLFTs amounts to imposing invariance under these two new local
moves.

Extending the program of refs.\ \refs{\BP,\FHK} to three dimensions, we
will give our construction an algebraic interpretation, and show how
to associate to each lattice theory an algebraic structure.  The
structure includes a multiplication operation, defined on faces, and a
comultiplication, defined on hinges (edges).  Reinterpreted in
algebraic terms, the condition of invariance under the hinge and cone
moves will be shown to imply that the multiplication and
comultiplication operations must satisfy the axioms of a Hopf algebra,
with an additional constraint on the antipode. (A brief review of Hopf
algebras is given in Appendix A.)  Furthermore, these constrained Hopf
algebras will be shown to be in one--to--one correspondence with 3D TLFTs.

In section 4, we will recover the original model of Ponzano and Regge
as a special case of our construction, with Hopf algebra based on the
group ring of SU(2). In addition, we will show that this model is
equivalent to SU(2) lattice gauge theory at zero coupling.

Finally, in section 5, we will offer some speculations on the
relationship of our TLFTs to other TFTs, and discuss generalizations
to higher dimensions.

\newsec{\bf Two--Dimensional Topological Lattice Field Theories}
\global\figno=1

Our study of topological lattice models begins in two dimensions.
Recently, two--dimensional topological lattice field theories were
shown to be in one--to--one correspondence with semisimple
associative algebras\
\refs{\BP,\FHK}. In this section we will review the definition and
classification of these models, from a point of view that will carry
over in subsequent sections to higher--dimensional TLFTs.

\subsec{Definitions}

By a two--dimensional lattice, we mean a collection of polygons with
edges identified pairwise. We will assume for now that the lattice is
without boundary, {\it i.e.,} that all edges are paired.
We will also assume that the lattice contains only triangular faces.
Both of these assumptions will be relaxed below.

A coloring of a lattice $L$ associates to each edge of each triangle
in $L$, an element $x$ of an index set $X$. A lattice field theory
(LFT) is defined by specifying a rule for assigning a weight to each
possible coloring $\{ x \}$. The partition function of the LFT is a
weighted sum over all possible colorings.

To be more specific, we will consider a class of theories in which the
weighting rule is defined locally on
individual triangles, and compute the overall weight by multiplying
the local weights of all the triangles together. Thus to each triangle
$\triangle$, with edges colored by $x$, $y$, and $z$, we assign a
complex number $C_{xyz}(\triangle )$. Since there is no canonical
ordering of the edges on $\triangle$, it is natural to impose cyclic
symmetry on the $C_{xyz}$;
\eqn\eCCYCLIC{
C_{xyz}=C_{yzx}=C_{zxy}.}
We will not, however, assume any relation between $C_{xyz}$ and
the orientation--reversed $C_{yxz}$.

The operation of identifying edges of adjacent triangles is effected
by means of a ``gluing operator'' on the space $X$, which we denote by
$g^{xy}$.  For the configuration depicted in \tfig\TwoA\ consisting of
two adjacent triangles, the associated weight is
\eqn\twotrianglesThree{
C_{xyu}~g^{uv}~C_{vzw}\equiv {C_{xy}}^{u}~C_{uzw}~,}
where as usual all repeated indices are summed over, and indices of
$C_{xyz}$ are raised by contracting with the gluing operator.
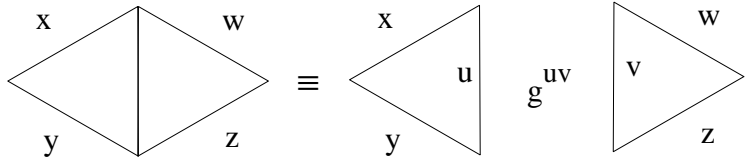
\ifig\TwoA{Two adjacent triangles.}{Fig-2-A}
With these conventions, the partition function of the LFT on a given
triangular lattice $L$ is
\eqn\eZTWO{
Z(L)\equiv\prod_{\triangle\in L}
\prod_{<uv>} C_{xyz}(\triangle )~g^{uv}~,}
where the second product is over all pairs of identified
edges.  For a lattice without boundary, the indices on all edges
should be contracted.

\subsec{Topological Invariance}

We would now like to find a set of conditions under which the
partition function \eZTWO\ is invariant under local changes of the
lattice $L$.

It is a theorem of Alexander that all triangulations of a topological
manifold can be generated by a set of local topology--preserving
lattice moves \ALEX.  One formulation of Alexander's
theorem\ \refs{\Mat,\Boulatov,\TV,\FHK}
says that just two basic moves are sufficient to generate all
topologically equivalent two--dimensional triangular lattices. The
first move, which flips an edge between two adjacent triangles, is
called the (2,2) move, because it converts two triangles into two new
triangles, as shown in
\tfig\TwoB a.
The second move, called the bubble move and drawn in \TwoB b,
collapses a pair of triangles with two edges in common to a single
edge.  According to the theorem, a lattice field theory, whose
partition function is invariant under both (2,2) and bubble moves, is
invariant under any local changes of triangulation, and thus depends
only on the topology of the underlying lattice. This is the defining
property of a topological lattice field theory.
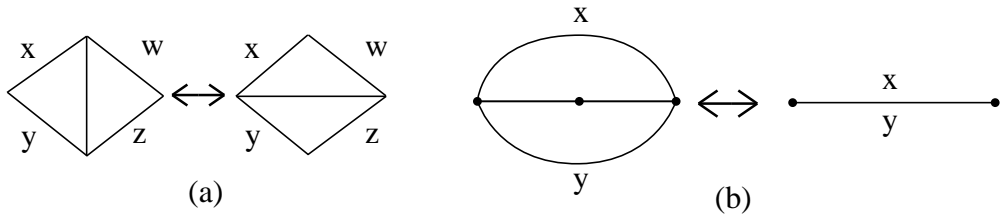
\ifig\TwoB{ (2,2) move and bubble move.}{Fig-2-B}
For the type of lattice field theory represented in eq.\ \eZTWO,
topological invariance has a simple expression in terms of the weights
$C_{xyz}$ and the gluing operator $g^{xy}$. Invariance under the (2,2)
move implies that
\eqn\twoduality{ {C_{xy}}^u {C_{uz}}^w={C_{xu}}^w {C_{yz}}^u, }
where we have raised the index $w$ for later convenience.  Likewise,
invariance under the bubble move is equivalent to
\eqn\twobub{
{C_{xu}}^v {C_{yv}}^u =g_{xy}~,}
where $g_{xy}$ is the matrix inverse of the gluing operator $g^{xy}$,
and will be called the ``metric.''

Equations \twoduality\ and \twobub\ are necessary and sufficient conditions
for an LFT defined by the data $\left( C_{xyz}, \, g^{xy} \right)$
to have topological invariance. As such, they may be taken as the
defining equations of a two--dimensional TLFT. Before proceeding with
their interpretation and solution, we would like to make the following
remarks.

\bigskip
\item{(I)}{(2,2) invariance (eq.\ \twoduality ) and bubble invariance
(eq.\ \twobub ) lead to the following equation
\eqn\etwocone{
{C_{xy}}^v\, {C_{vu}}^u =g_{xy}~.}
This equation guarantees invariance under the ``cone move,''
which transforms the cone--like configuration of two triangles
in\tfig\TWOunitone a into a single edge in \TWOunitone b.
(For the sake of clarity,  two additional triangles $C_{xab}$ and $C_{ycd}$
are attached to the edges $x$ and $y$ in \TWOunitone c.)
\tfig\TWOunittwo\ gives a graphical proof of eq.\ \etwocone , and
also shows that, conversely,
the cone move and the (2,2) move generate the bubble move; hence,
the cone and (2,2) moves are an alternative set of moves that generate
all topology--preserving deformations.}
\ifig\TWOunitone{Cone move.}{Fig-2-unitone}
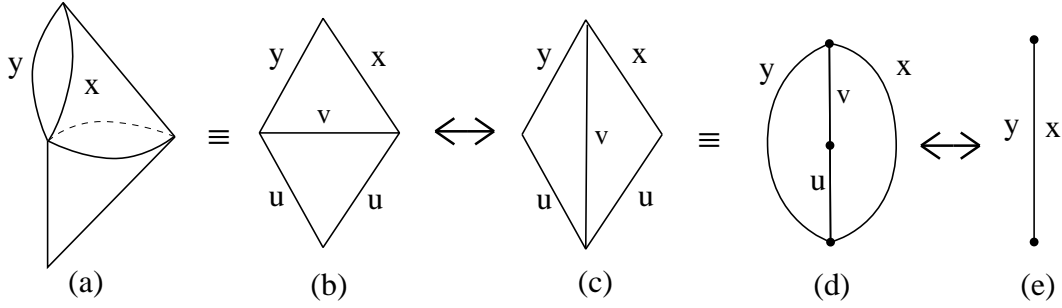
\ifig\TWOunittwo{Cone move from bubble move.}{Fig-2-unittwo}

\item{(II)}{The (2,2) and bubble moves can actually be interpreted
as lattice analogues
of local metric deformations in continuum. In fact, the (2,2) move, which
preserves locally the number of triangles, is a lattice analogue of an
area--preserving diffeomorphism. The bubble move decreases the number
of triangles in a local region of the lattice; it is like a Weyl
transformation.  In  continuum, it is well--known that
area--preserving diffeomorphisms and Weyl transformations generate all
local deformations of metric. The corresponding statement here is
that (2,2) and bubble moves generate all local topology--preserving
deformations of a triangular lattice. It is also interesting to study
lattice theories invariant under only (2,2) or bubble transformations.
For example, two--dimensional lattice gauge theory can be shown to be
invariant under the (2,2) move, corresponding to its invariance under
area--preserving diffeomorphisms in continuum \refs{\Migdal,\Wittenfour}.}

\item{(III)}{Up to now, we have considered only triangular lattices.
With little additional effort, we can relax this assumption, and extend our
arguments to lattices with non--triangular as well as triangular faces.
To define a general LFT
on such a lattice, we need to introduce an infinite set of new weights
$C_{x_1 \ldots x_n}$, in order to account for the
contribution of an arbitrary $n$--gon.  Now the topology--preserving
moves that we have defined so far are sufficient to relate all
topologically equivalent triangular lattices; however, to obtain all
topologically equivalent lattices with general polygonal faces as
well, we must consider additional moves that generate subdivisions of
polygons into triangles.  After imposing subdivision invariance on
the generalized weights $C_{x_1 \ldots x_n}$, we find that they
 are completely
determined by the old, triangular weights, since $n$--gons can always
be decomposed into triangles; for the decomposition shown in\tfig\kgonFIG ,
we obtain
\eqn\eCCC{
C_{x_1 \ldots x_n}=C_{a_1x_1b_1}\, g^{b_1 a_2}\, C_{a_2x_2b_2}\,g^{b_2 a_3}
\cdots C_{a_nx_nb_{n}}\,g^{b_n a_1}~.}
Topological invariance guarantees that the generalized weights so
obtained are well--defined, that is, independent of the triangular
decomposition chosen.  Because all higher weights are completely
determined by the original triangular weights, no new topological
field theories are obtained by considering arbitrary polygonal
lattices, and 2D TLFTs on arbitrary lattices are in one--to--one
correspondence with solutions to the (2,2) and bubble equations,
eqs.\ \twoduality\ and \twobub. }
\ifig\kgonFIG{A decomposition of an $n$--gon.}{Fig-4-one}

\subsec{Algebraic Structure}

We will now explain how a solution $\left( C_{xyz}, \, g^{xy}\right)$ to
the (2,2) and bubble equations, besides defining a TLFT, also determines an
algebraic structure.  Namely, we will review that given any
two--dimensional TLFT, we can construct a semisimple associative
algebra \refs{\BP,\FHK}.

The algebra in question is defined on the the vector space $A$ over
$\C$ generated by the index set $X$:
\eqn\eRALGEBRA{
A={\bigoplus_{x\in X}} ~\C \phi_x ~.}
We introduce a multiplication operation on $A$, by means of the
${C_{xy}}^z\equiv C_{xyz'}g^{z'z}$:
\eqn\eMULT{
\phi_x \cdot \phi_y\equiv {C_{xy}}^z\phi_z ~.}
Then it is easy to see that eq.\ \twoduality \ -- invariance under the
(2,2) move -- is equivalent to the associativity of the
multiplication \refs{\BP, \FHK}:
\eqn\eASSOC{
(\phi_x \cdot\phi_y)\cdot\phi_z=\phi_x\cdot(\phi_y\cdot\phi_z)~.}

Furthermore, in the bubble equation \twobub, we implicitly assume that the
inverse of $g_{xy}$ exists, which is the gluing operator $g^{xy}$. Note here
that the nondegeneracy of $g_{xy}$ is equivalent to the condition that
$A$ be semisimple \FHK. Thus, given a topologically invariant lattice field
theory, defined by data $C_{xyz}$ and $g^{xy}$ satisfying \twoduality\
and \twobub, we can always construct a semisimple associative algebra.

Conversely,  a TLFT can be constructed from a given
 semisimple associative algebra on a vector space
$A$ with structure constants ${C_{xy}}^z$ \refs{\BP,\FHK}. By applying
the quadratic form $g_{xy}={C_{xu}}^v {C_{yv}}^u$ to
the structure constants, we construct a set of triangular weights
$C_{xyz}\equiv {C_{xy}}^{z'}g_{z'z}$, which can be shown to be cyclic
by using the associativity condition \twoduality.  The semisimple
condition implies that the matrix inverse $g^{xy}$ of $g_{xy}$ exists;
this is our gluing operator. Since the LFT data $C_{xyz}$ and $g^{xy}$
satisfy eqs.\ \twoduality\ and \twobub\ by assumption, they
automatically define a TLFT.

In summary, we have shown
\medskip

\noindent {\bf Theorem 2.1.}\ \refs{\BP,\FHK} {\it Two-dimensional
topological lattice field theories  are in
one--to--one correspondence with semisimple associative algebras.}
\bigskip

We close this subsection with a remark on the algebraic interpretation
of the ``cone equation'' \etwocone : in a 2D LFT whose gluing operator
has its inverse, invariance under the cone move is equivalent to the
existence of a unit element $1_A$ in the corresponding algebra $A$.
In fact, by rewriting eq.\ \etwocone\ in the form
\eqn\UNITDELTA{
u^x\,{C_{xy}}^z={\delta_{y}}^z =u^x\, {C_{yx}}^z~}
with $u^x\equiv g^{xy}\, {C_{yu}}^u$,
we obtain an equation equivalent to the defining property of a unit element
$1_A\equiv u^x\,\phi_x$ of $A$:
\eqn\UNIT{1_A\cdot\phi=\phi=\phi\cdot 1_A~,~~~~~~ \forall\,\phi\,
\in\, A~.}
This result corresponds to the mathematical fact that in any semisimple
associative algebra, we can uniquely introduce a unit element $1_A$ of
the above form (see Theorem A.1).

\subsec{Physical Observables}

Our main concern up to now has been the definition of partition
functions with topological invariance. We will now describe how
to construct physical operators and states.

A natural operator to consider is the operator $\O_{xyz}(\triangle)$
that fixes the colors $x$, $y$, and $z$ on the edges of a single triangle
$\triangle$.  The formal definition of this operator,
which is conjugate to $C^{xyz}(\triangle )$, is
\eqn\eOP{
\O_{xyz}(\triangle) =
{\delta \over \delta C^{xyz}(\triangle )}~,}
so that  correlation functions are given by
\eqn\eCORR{
\langle \O_{x_1 y_1 z_1}({\triangle_1})~\cdots~
	\O_{x_k y_k z_k}({\triangle_k})
    \rangle ~=~{\delta \over \delta C^{x_1 y_1 z_1}(\triangle_1)}
     ~\cdots~{\delta \over \delta C^{x_k y_k z_k}(\triangle_k)}~Z~.}
The insertion of an operator $\O_{xyz}(\triangle)$ causes the sum over
colorings of the lattice to be restricted to those colorings that have
the fixed
values $x$, $y$, and $z$ on the edges of $\triangle$, and removes the
weighting factor $C^{xyz}$ of the triangle $\triangle$ from the partition
function \eZTWO .
$\O_{xyz}(\triangle)$ thus effectively punches a triangular hole in the
lattice with boundary $\triangle$
and boundary values $x$, $y$, and $z$.  We can further construct an
operator that creates  an arbitrary polygonal
boundary by gluing triangular operators $\O_{xyz}$ together and
summing over glued edges. For example, if $\triangle_1$
and $\triangle_2$ are two adjacent triangles making up a
quadrilateral $\square$ with edges $x$, $y$, $z$, and $w$, then
we can define $\O_{xyzw}(\,\square\,)$ as
\eqn\eOOO{
\O_{xyzw}(\,\square\,)\equiv {\delta\over \delta C^{xyzw}}=
{\O_{xy}}^u(\triangle_1)\, O_{uzw}(\triangle_2)~.}

The colorings of a fixed boundary created by the operators
$O_{x_1x_2\ldots x_n}$ are naturally regarded as forming a Hilbert space of
states.  In order to determine which of these states are physical, we
introduce the following ``Hamiltonian formalism'' (following ref.\
\ATI ).\foot{An alternative approach is given in ref.\ \FHK.} First,
we give a more precise definition of the Hilbert space ${\cal H}_B$
(not necessarily physical) as the module freely generated by
all possible colorings $C$ of a ``triangulated'' one-dimensional closed
manifold $B$.  Thus, a wave function in ${\cal H}_B$ is a
complex--valued function of a coloring on $B$; ${\cal H}_{B}=\{
\Phi_{B}(C)\}$.  We then define the ``time''--evolution kernel
$P_{B',B}(C',\,C)$, which maps a state $\Phi_{B}(C)$ in ${\cal H}_{B}$
to a state ${\Phi'}_{B'}(C')$ in ${\cal H}_{B'}$, as the cylinder with
boundary $B'\cup (-{B})$.
Here, $-{B}$ denotes the orientation
reverse of $B$. In a topological theory, the cylinder
$P_{B',B}(C',\,C)$ should be independent of its triangulation;
in particular it should obey the composition law
\eqn\eCOMP{
P_{B'',B}(C'',\,C) = \sum_{C'}\, P_{B'',B'}(C'',\,C')\,
P_{B',B}(C',\,C)}
for arbitrary intermediate boundary $B'$.
Taking $B=B'=B''$ in \eCOMP, we see that $P_{B,B}^2=P_{B,B}$, \ie, the
cylinder $P_{B,B}$ is a projection operator.
Physical states $\Psi_{B}(C)$, which appear on the intermediate slice $B'=B$,
can be characterized by their invariance under ``time''--evolution by
$P_{B,B}$:
\eqn\eq{
\Psi_{B}(C')= \sum_{C}\, P_{B,B}(C',\,C)\, \Psi_{B}(C)~.}
$P_{B,B}$ is thus the projection operator of ${\cal
H}_{B}$ onto the physical Hilbert space ${\cal H}_{B}^{\rm phys}$, and
is the propagator of physical states.

For example, let us determine the physical states on a lattice with a
boundary consisting of a single edge, with coloring $x$.  The
projection operator should correspond to a cylinder with two
one--gonal holes, with colorings $x$ and $y$. Such a cylinder can be
triangulated as in \tfig\TwoD , and from this triangulation we
determine that the propagator is
\eqn\eONEGON{
{P_x}^y={C_{xu}}^v\, {C_{v}}^{yu}~.}
This is precisely the projection operator onto one--gonal physical
states that was found in ref.\ \FHK\ (where it is called ${\eta_x}^y$).
There, it was further shown that
${P_x}^y$ is algebraically a projection operator from the semisimple
associative algebra $A$ onto its
center $Z(A)$, and that there is a one--to--one correspondence of
elements of $Z(A)$ with physical operators.
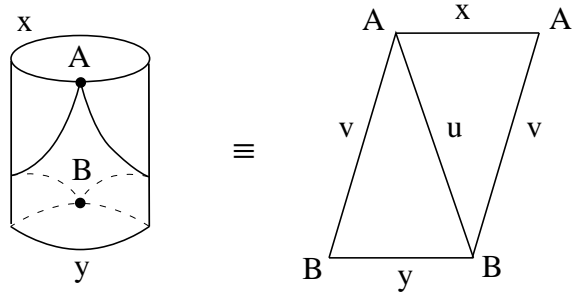
\ifig\TwoD{A triangulation of a cylinder with two one--gonal holes, $x$
and $y$.}{Fig-2-D}

Besides the propagator, we can define an interaction vertex for physical states
in terms of the 3--point function of 3 separated one--gons
 shown in \tfig\TwoE :
\eqn\eTHREEPT{
N_{xyz}\equiv{P_x}^p {P_y}^q {P_z}^r\,C_{pqr}
={P_x}^p {P_y}^q {P_z}^r\,C_{qpr}=N_{yxz}~.}
Eq.\ \eTHREEPT\ implies that
$N_{xyz}$ is the totally symmetric projection of $C_{xyz}$ to
$Z(R)$. These data $P$ and $N$ are sufficient to calculate all correlation
functions of physical operators of arbitrary genus.
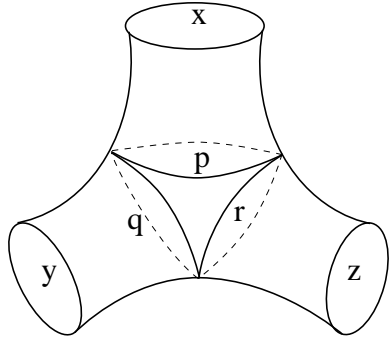
\ifig\TwoE{Three-point function of $\O_x$, $\O_y$, and $\O_z$.}{Fig-2-E}

\subsec{Example: $A=\C [G]$}

A simple example of an associative algebra is provided by group ring
$A=\C [G] = \oplus_{x\in G}\, \C \,\phi_x$, with multiplication
inherited from the group $G$, {$\phi_x \cdot \phi_y=\phi_{xy}$.}
Let $\int {\rm d}x$ be its normalized Haar measure:
\eqn\eHAAR{\eqalign{
&\int {\rm d}~(gxh)=\int {\rm d}x~~~~~~~~~~(\forall ~g,h~\in~G)\cr
&\int {\rm d}x ~1 =1~.\cr}}
Associated with this Haar measure is the left--right invariant
$\delta$-function $\delta(x,y)$:
\eqn\eDELTA{\eqalign{
&\int {\rm d}y ~\delta(x,y)~f(y)= f(x)\cr
&\delta (gxh,~gyh)=\delta(x,y)~,~~~~~~~~~~(\forall ~x,~y,~g,h~\in~G)~.\cr}}
For a finite group $G$, we define the measure and $\delta$--function by
\foot{In ref.\ \FHK , $\delta_{x,y}$
is denoted by $\delta(x,y)$.}
\eqn\eq{\eqalign{
&\int {\rm d}x \equiv {1\over{\gn}}\sum_{x\in G}\cr
&\delta(x,y)\equiv \gn~ \delta_{x,y}~.\cr}}

The structure constants for $A=\C [G]$ are thus
\eqn\eSTRCON{
{C_{xy}}^z=\delta(xy,z)~.}
Following eq.\ \twobub\ we have
\eqn\eGC{\eqalign{
g_{xy}&={C_{xu}}^v{C_{yv}}^u= \int {\rm d}u\,{\rm d}v\, \delta(xu,v)\,
    \delta(yv,u) = \delta(x,y^{-1})\cr
C_{xyz}&={C_{xy}}^u \,g_{uz} = \int {\rm d}u\,\delta(xy,u)\,\delta(u,z^{-1})
      =\delta(xyz,1_G)~.  \cr}}
Now we can readily
calculate the two-- and three--point functions of one--gons on
the sphere, using the same triangulations as in eqs.\ \eONEGON\ and
\eTHREEPT:
\eqn\eTWOPTONE{
\eqalign{
P_{xy}&={C_{xu}}^v {C_{vy}}^u= \int {\rm d}u\, {\rm d}v\,
\delta(xu,v)\,\delta(vy,u)=\int {\rm d}v\,\delta(x,vy^{-1}v^{-1})=
\delta_{[x],[y^{-1}]}\cr
&= \sum_j \chi_j(x) \chi_j(y)\cr
N_{xyz}&=\int {\rm d}x'\,{\rm d}y'\,{\rm d}z'
P_{x}^{x'} P_{y}^{y'} P_{z}^{z'}\,\delta(x'y'z',1_G)
= \sum_j {\chi_j(x) \chi_j(y) \chi_j(z)\over {d_j}}~,\cr}}
where $[x]$ denotes the conjugacy class
of $x$, and, $\chi_j$ and $d_j$ are, respectively,
 the group character and the dimension of an irreducible representation $j$.
By iteration, the correlation function for $n$ one--gons on genus $g$
can be computed to be \refs{\Wittenfour,\BP,\FHK}
\eqn\eNPOINT{
\langle \o{x_1} \o{x_2}\cdots \o{x_n} \rangle_g
=\sum_j {\chi_j(x_1) \chi_j(x_2) \cdots \chi_j(x_n)
\over{d_j^{n+2g-2} }}~.}

Arbitrary correlation functions of the $p$--gon operators
$\O_{x_1\cdots x_p}$ can easily be computed as well, and are no
more complicated than correlation functions of one--gons.
In fact, we only have to connect a cylinder with two boundary loops, one of
which is one--gonal, and the other is $p$-gonal.
For example, in the case of $p=2$,
we attach one edge of a triangle, with free edges $y$ and $z$ forming
a two--gon,
to a cylinder with one--gon ends, as in \tfig\TwoF . The result is simply
\eqn\eONETWO{
\langle \O_{x}\O_{yz} \rangle_{g=0} = {P_x}^{x'} C_{x'yz} = P_{x,yz}
=\sum_j\, \chi_j(x)\, \chi_j(yz)~.}
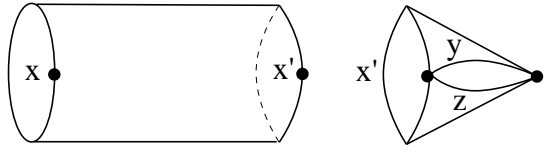
\ifig\TwoF{Two point-function of $\O_x$ and $\O_{yz}$.}{Fig-2-F}
\noindent    Similarly, we have the following expression for arbitrary $p_i$:
\eqn\eONEN{
\langle \O_{x}\O_{y_1\cdots y_{p}} \rangle_{g=0}
=P_{x,y_1\cdots y_{p}}
=\sum_j \chi_j(x) \chi_j(y_1 \cdots y_{p})~.}
By attaching two--point functions of the form \eONEN\ to one--gonal
correlation functions such as \eNPOINT, we can build any correlation
function of $p$--gonal operators. The result is of the same form as
eq.\ \eNPOINT, but with $x_i$ replaced by the product of group
elements around the $i$th polygon:
\eqn\eGENCF{\eqalign{
\langle &\o{x_{(1,1)},\ldots,x_{(1,p_1)}} \cdots
\o{x_{(n,1)},\ldots,x_{(n,p_n)}}\rangle_g\cr
&=\sum_j {\chi_j(x_{(1,1)}\cdots x_{(1,p_1)})
\cdots \chi_j\left(x_{(n,1)}\cdots x_{(n,p_n)}\right)
\over{d_j^{n+2g-2} }}~,\cr}}
where $x_{(i,j)}$ is the group element on the $j$th edge of the $i$th polygon.

\subsec{Dual Formulation}

There is an equivalent approach to  constructing TLFTs, in which
interactions take place at vertices instead of on triangles.
For simplicity, but
ultimately without loss of generality, we will suppose that all
vertices are trivalent.  Then given a lattice whose links form a
trivalent graph, we can
construct a lattice field theory by gluing all the vertices together.
The weight $\D^{xyz}$ of a vertex will depend on the colorings of the
three lines emanating from it, and vertices will be glued using a
two--indexed operator $h_{xy}$. Similarly to eq.\ \eZTWO , the partition
function of this theory will be the product of all vertex weights
$\D^{xyz}$, with all indicies contracted using $h_{xy}$.

Of course, this lattice model is completely dual to our previous
construction, and is in all respects mathematically equivalent, but it
may be instructive to review the axioms of topological invariance in
this framework. For this purpose, we may replace any of our previous
pictures of triangles glued together along links, by dual graphs, in
which dual vertices are connected by dual links. Then (2,2) invariance
(eq.\ \twoduality) of the original lattice theory is now expressed as
\eqn\eDUALtwoduality{
{\D_x}^{yu}\, {\D_u}^{zw}={\D_x}^{uw}\, {\D_u}^{yz}} in this dual
picture (see\tfig\TwoG a).  Furthermore, we can consider the ``dual
bubble move,'' that is, the move in the dual lattice corresponding to
the bubble move in the original lattice,
\eqn\eDUALbubble{
h^{xy}={\D_u}^{vx}\,{\D_v}^{uy},}
 from whose graphical expression (\TwoG b) the bubble move derives its name.
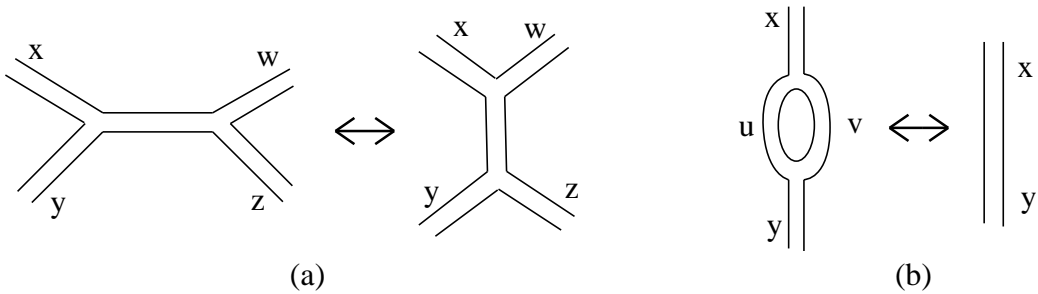
\ifig\TwoG{Dual diagrams of \TwoB .}{Fig-2-G}
We may also consider lattice theories containing both $C_{xyz}$'s and
$\Delta^{xyz}$'s. In two dimensions, this just gives two uncoupled
lattice models of the type we have already discussed. However, in the
following section, an elaboration of this idea will form the basis for
our construction of topological lattice theories in three dimensions.

\newsec{\bf Construction of TLFTs in Three Dimensions}
\global\figno=1

In the previous section, we described a one--to--one correspondence
between a class of 2D topological lattice field theories and
semisimple associative algebras.  Here we will show that in
three dimensions there is a similar sort of correspondence between a
class of TLFTs and a class of Hopf algebras.  We begin by defining the
class of lattice field theories (LFTs) of interest and establishing a
diagrammatic representation for operations within these theories
(subsec.\ 3.1). We then show that the data of these LFTs can be used
to define algebra and coalgebra structures (subsec.\ 3.2).  Our main
result will be that these structures combine to form a special type of
Hopf algebra if and only if the theory is invariant under
arbitrary topology--preserving deformations of
the lattice geometry (subsec.\ 3.3).

\subsec{Definitions}

In this subsection, we define a general class of lattice field
theories in three dimensions, valid for an arbitrary lattice.

One possible way of describing a three--dimensional lattice $L$ is as
a collection of oriented polyhedra with faces glued pairwise.  Such a
lattice can be colored by associating to each of its edges an element
$x$ of an index set $X$. To define a local lattice field theory on
$L$, we specify a rule that determines a local weight for each
polyhedron, as a function of the colorings of its edges. The total
weight of a given coloring is then computed by taking the product of
the local weights of all polyhedral cells in $L$; in turn, the
partition function is defined to be the sum of these total weights
over all allowed colorings.

This approach (especially for tetrahedral lattices) has formed the
basis for most previous constructions of TLFTs in three
dimensions \refs{\RP,\TV,\OS,\DJN}. However, we will find that
the following alternative approach is better suited to our purpose of
establishing a correspondence between TLFTs and algebraic structures.

We define a lattice $L$ to be a collection of polygonal faces, with
edges glued together -- not necessarily pairwise -- along
one--dimensional hinges.  We again color the lattice by associating to
each edge of each face a color $x \in X$.\foot{Other coloring and
weighting schemes are discussed in Sec.\ 5, as well as in ref.\ \FG.}
As in two dimensions, the coloring of the boundary of each face
determines a local weight.  But in contrast to the two--dimensional
case, in three dimensions it is possible for the edges of more than
two faces to meet, as in \tfig\fGEN a.  We call a meeting of three or more
edges a ``hinge;'' more precisely, a hinge is an open neighborhood of
the line along which the faces meet. Hinges thus have edges that can be
colored, and can be pictured as consisting of several infinitesimally
thin strips emerging from a central line.  For example, a
 four--valent hinge is contained in \fGEN b.
Because hinges, as well as faces, have edges that need to be colored
and glued, a weighting rule for hinges must also be specified. This
will be a function of the colors on the edges of the hinge, to be
called the ``hinge weight.''
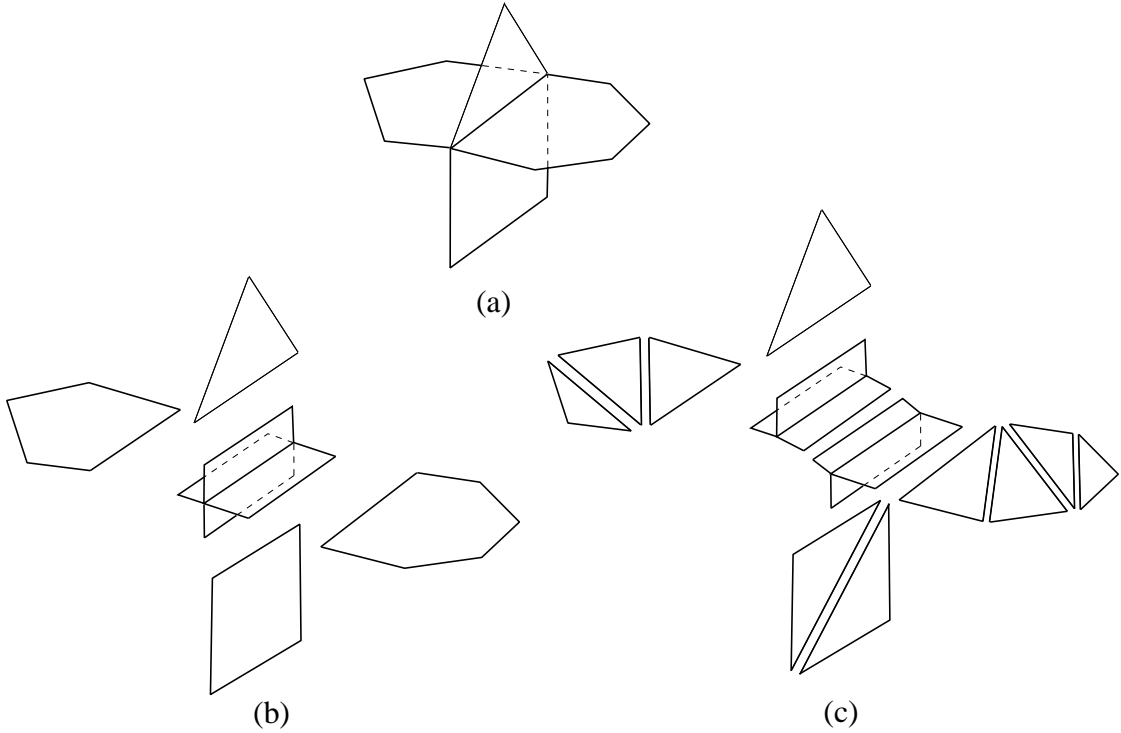
\ifig\fGEN{The decomposition of a 3D lattice into faces and
hinges.}{Fig-3-general}

The weighting rules for faces and hinges that we will consider will be
of a particular type. Namely, we will further decompose, respectively,
 each polygonal
face and each multivalent hinge into triangles and trivalent
hinges (3--hinges),  as depicted in \fGEN c, and
demand that our weighting rules be invariant under subdivision.
This initial restriction,
which is automatically satisfied in a topologically invariant theory,
will not affect our conclusions with regard to 3D TLFTs, yet will
greatly simplify the form of the weighting rules.
Thus, in order to determine a complete set of weights for arbitrary
faces and hinges, it will suffice to specify weights only for
triangles and 3--hinges, and ``gluing rules'' that tell us how to
connect triangles to triangles (so as to compute the weight of
an arbitrary polygonal face in terms of triangular weights),
and 3--hinges to 3--hinges (so as to compute the weight of an
arbitrary multivalent hinge).

Besides weights for faces and hinges, we also have to specify how to
connect faces to hinges, so as to form a 3D lattice.  Since
faces and hinges are already decomposed into triangles and 3--hinges,
we only need to introduce additional gluing rules that tell us how to
connect triangles to
3--hinges. Then, given this collection of triangles and 3--hinges,
we define
the partition function of our LFT to be the product of all triangle
 weights and 3--hinge
weights, with color indices contracted according to the gluing rules.

\bigskip

We first define a weighting rule for triangular faces.  To each
triangle, whose three edges are colored by $x$, $y$, and $z\in X$, we
associate a complex number $C_{xyz}$. The weighting rule for faces is
thus a map $C: X\times X\times X\to \C$. As in two dimensions, the
symmetry of the triangle imposes cyclic symmetry on
$C_{xyz}=C_{zxy}=C_{yzx}$.  In order to keep track of the orientation
of each triangle, we now draw arrows around its edges, as in \tfig\fA.
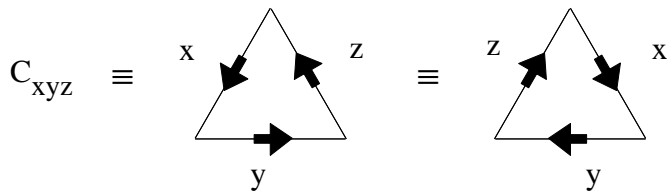
\ifig\fA{Weight of a triangle $C_{xyz}$.}{Fig-3-one}

Similarly, we define a weighting rule for
3--hinges.  A 3--hinge consists of three infinitesimally thin
strips, as shown in {\tfig\fE}a; its three outward edges are colored
and will eventually be glued to faces (or to other hinges).  For
artistic convenience, we represent this hinge in {\fE}b, as a
diagram consisting of three parallel dotted lines with aligned arrows,
labeled $x$, $y$, and $z$.  We assign a local weight $\Delta^{xyz}$ to
the hinge, if the three edges meeting on the hinge have colors $x$,
$y$, and $z$. When the arrows on the parallel lines point upwards, our
convention will be to orient $x$, $y$, and $z$ counterclockwise,
viewed from above. Because none of the three faces is preferred,
cyclic symmetry is also imposed:
$\Delta^{xyz}=\Delta^{yzx}=\Delta^{zxy}$.
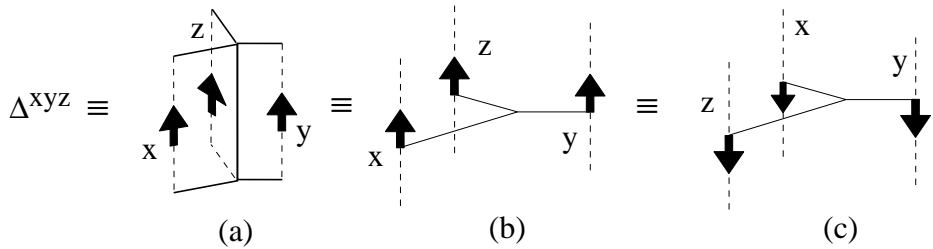
\ifig\fE{Graphical representation of $\Delta^{xyz}$.}{Fig-3-five}

\bigskip

Next, we will explain  how to glue triangles to triangles and 3--hinges
to 3--hinges. {\phantom{\tfig\fB}}

To glue two edges of two triangular faces together, with colors $x$
and $y$, we introduce a ``face--gluing operator'' $g^{xy}$. That is,
we use $g^{xy}$ to contract the indices of the triangular weights that
correspond to the edges to be glued.  We represent the operator
$g^{xy}$ by two parallel dotted lines, with arrows pointed in opposite
directions, as in \fB.  In all of our diagrams, we will use
dotted lines to represent raised indices and solid lines for lowered
indices, with the rule that solid lines may only be glued to dotted
lines, and then only if the arrows on the two lines have the same
orientation.  This rule is essentially the usual summation convention
for repeated indices.  For example, for the configuration
of two triangles shown in\tfig\fC , the corresponding local weight is
$C_{xyu}\,g^{uv}\,C_{vzw}\equiv {C_{xy}}^u\, C_{uzw}$.
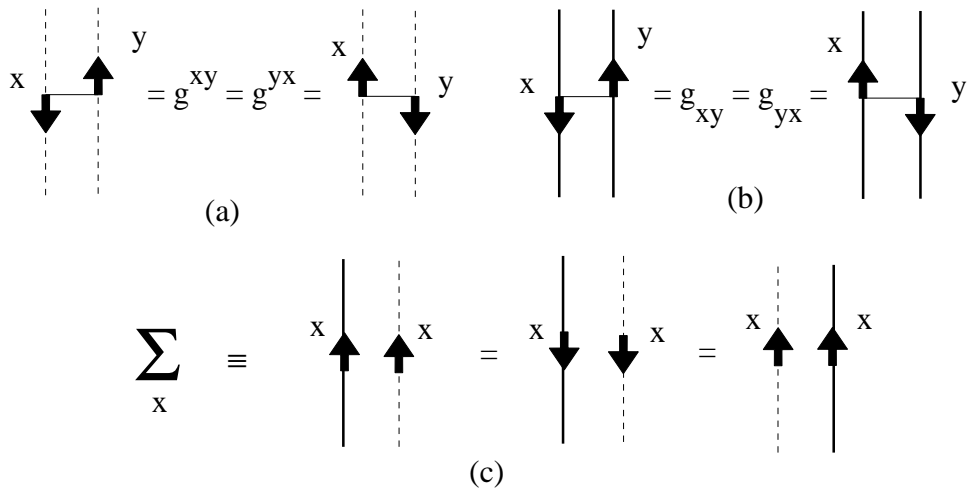
\ifig\fB{Definitions of $g_{xy}$, $g^{xy}$ and summation.}{Fig-3-two}

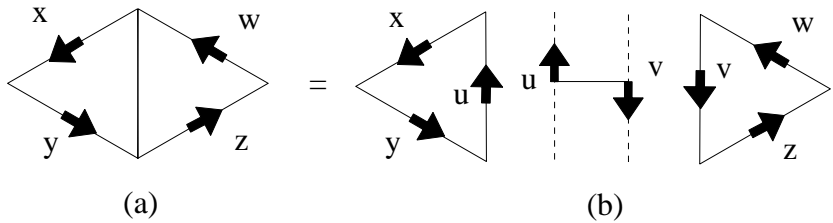
\ifig\fC{Two triangles joined by $g^{uv}$.}{Fig-3-three}

Given a set of triangle weights $C_{xyz}$ and gluing operators
$g^{xy}$, we can now compute the weight of an arbitrary polygonal face,
all of whose external edges are oriented in the same direction ({\it
i.e.,} clockwise or counterclockwise), by decomposing the polygon into
a collection of triangles, similarly oriented. The weight of the
polygonal face is defined to be the product of the local weights of
the triangles, with all internal edge indices contracted using the
face--gluing operator $g^{xy}$.  For example, the local weight of the
$n$-gonal face shown in \tfig\fI\ is
\eqn\eNFACE{\eqalign{
C_{x_1x_2...x_n}&\equiv C_{a_1x_1a_2}~g^{a_2a_2^{'}}~
C_{a_2^{'}x_2a_3}~g^{a_3a_3^{'}}~...~g^{a_na_n^{'}}~
C_{a_n^{'}x_na_1^{'}}~g^{a_1^{'}a_1}\cr
&=C_{a_1x_1}^{{\phantom{a_1x_1}a_2}}~
C_{a_2x_2}^{{\phantom{a_2x_2}a_3}}~...~C_{a_nx_n}^{{\phantom{a_nx_n}a_1}}~.
\cr}}
\ifig\fI{A decomposition of a polygonal face.}{Fig-3-nine}

For consistency, the definition of polygonal face weights in terms of
triangular constituents must be independent of the triangular
decomposition chosen.
This will be the case if and only if we impose the same two
constraints on $C_{xyz}$ and $g^{xy}$ that guaranteed topological
invariance of the two--dimensional lattice field theory in sec.\ 2,
namely, invariance under the (2,2) move and the bubble move, eqs.\
\twoduality\ and \twobub:
\eqn\eCABCD{\eqalign{
{C_{xy}}^u {C_{uz}}^w&={C_{xu}}^w {C_{yz}}^u\cr
{C_{xu}}^v\, {C_{yv}}^u &=g_{xy}~,\cr}}
where $g_{xy}$ is the inverse of $g^{xy}$ and will be called the ``metric.''

{\phantom{\tfig\fF}}
\bigskip

To glue 3--hinges to 3--hinges, we introduce a
``hinge--gluing operator'' $h_{xy}$, whose function is to lower the
indices of $\Delta^{xyz}$.
$h_{xy}$ is represented graphically by a pair of parallel solid lines
with the arrows in the same direction {(\fF )}, and always joins
hinges with arrows aligned.  It is important to remember that just as
$g^{xy}$ is used to raise the indices of $C_{xyz}$ only, so $h_{xy}$
can only lower the indices of $\Delta^{xyz}$ (and not, \eg, $C^{xyz}$).

Now we can assign weights to multivalent hinges, with a prescription similar
to the prescription for polygonal faces.
Namely, just as an $n$--sided polygon can be decomposed into
triangles, an $n$--valent hinge can be decomposed into a collection of
3--hinges. The weight of an $n$--hinge is equal to the product of the
weights of its constituent 3--hinges, with internal color indices
contracted using the hinge--gluing operator $h_{xy}$.
For example, to compute the weight of the 4--hinge $\Delta^{xyzw}$ shown
in {\tfig\fJ}a, we
can decompose it into two 3--hinges, contracted with a hinge--gluing
operator, as in {\fJ}b:
\eqn\eFOURHINGE{
\Delta^{xyzw}\equiv \Delta^{xyu}h_{uv}\Delta^{vzw}\equiv \Delta^{xyu}
\Delta_u^{{\phantom{u}}zw}~. }
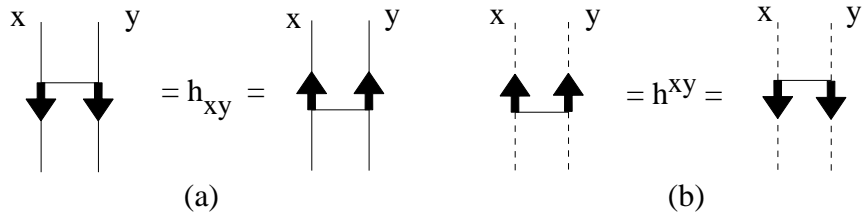
\ifig\fF{Graphical representation of $h_{xy}$ and $h^{xy}$.}{Fig-3-six}
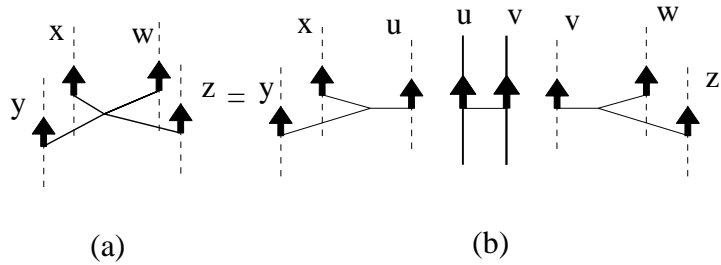
\ifig\fJ{Definition of a 4--hinge.}{Fig-3-ten}
\noindent Similarly, for
the $n$--hinge shown in \tfig\fK , the corresponding local weight is
defined as
\eqn\eNHINGE{
\Delta^{x_1x_2...x_n}\equiv{\Delta_{a_1}}^{x_1a_2}
{\Delta_{a_2}}^{x_2a_3}\,\ldots\,{\Delta_{a_n}}^{x_na_1}~.}
\ifig\fK{An $n$--hinge.}{Fig-3-eleven}

As was the case for face weights, consistency requires that our
definition of $n$--hinge weights be independent of the decomposition
into 3--hinges chosen. Since, on the plane perpendicular to the $n$--hinge,
this requirement is equivalent to the 2D topological invariance that
we imposed on trivalent  graphs in subsec.\ 2.6, we only need to
require eqs.\ \eDUALtwoduality\ and \eDUALbubble ; that is,
{\phantom{ {\eqnn\eINDEP}  {\eqnn\eDBUBBLE} }}
$$\eqalignno{
&\Delta_x^{{\phantom{x}}yu}\, \Delta_u^{{\phantom{u}}zw}
= \Delta_x^{{\phantom{x}}uw} \Delta_u^{{\phantom{u}}yz} &\eINDEP \cr
&\Delta^{{\phantom{v}}ux}_{v}\Delta^{{\phantom{u}}vy}_{u}=
     h^{xy}~, &\eDBUBBLE  \cr}  $$
where $h^{xy}$ is the inverse of $h_{xy}$ and will be called the ``cometric.''
It may  be useful to recall the geometrical meaning of these equations.
Eq.\ \eINDEP\ insures that the definition of $\D^{xyzw}$ is independent of
whether we decompose the {4--hinge} as in\tfig\fJBB a or b,
while eq.\ \eDBUBBLE\
guarantees that we can collapse a hinge loop, as depicted in\tfig\fL .
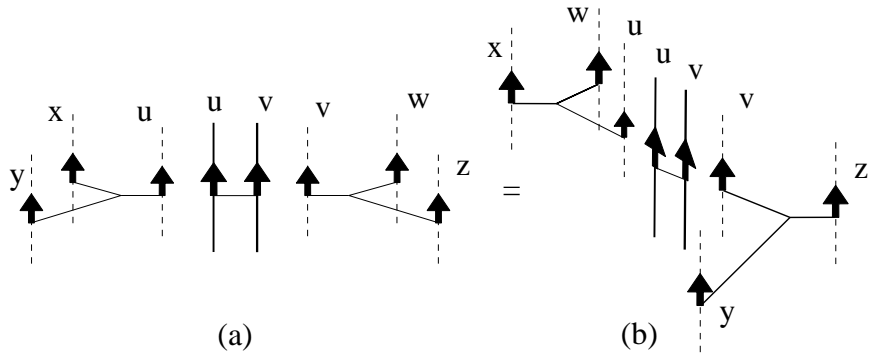
\ifig\fJBB{Two different decompositions of a 4--hinge.}{Fig-3-tenB}
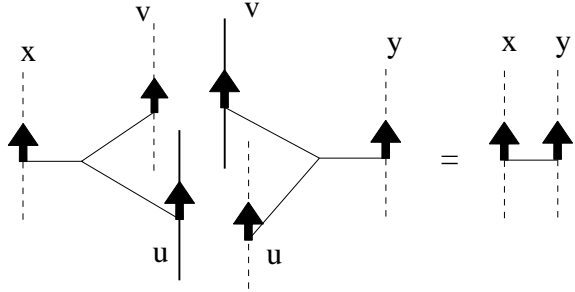
\ifig\fL{Reduction of a hinge-loop.}{Fig-3-twelve}

\bigskip

Having described how to glue triangles to triangles and 3--hinges to
3--hinges,  we
still need to explain how to glue triangles to 3--hinges.  There are
two separate cases to consider, depending on whether the arrows of the
face edge and the hinge edge to be joined are in the same direction
or in opposite directions.    {\phantom{\tfig\fD }}
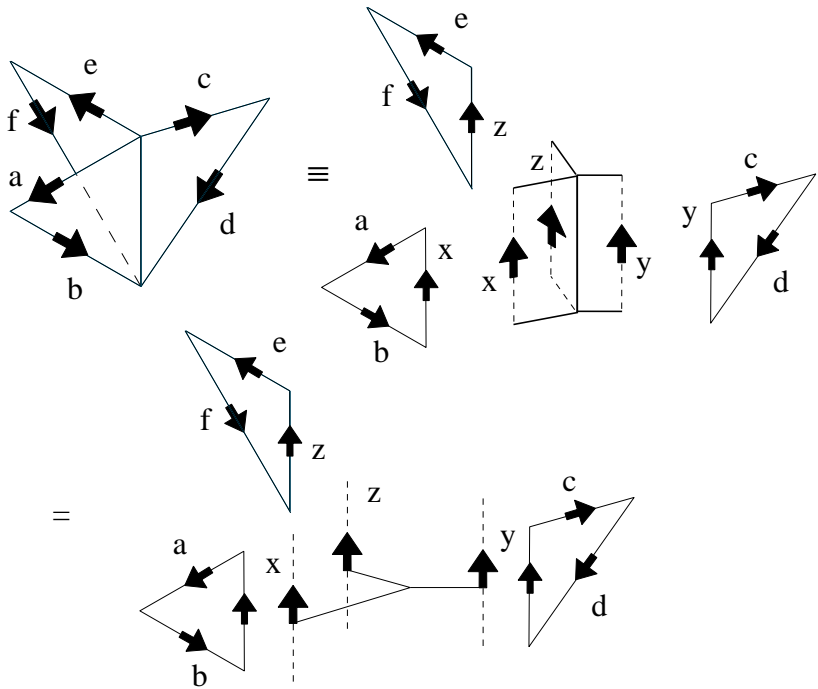
\ifig\fD{Three triangles joined along a hinge.}{Fig-3-four}
For the former case, our gluing rule will be simply to contract the
corresponding lower face index and upper hinge index.\foot{In principle,
we could generalize this particular gluing rule by
introducing a new gluing operator ${Q^x}_y$, to insert between any
3--hinge and triangle indices to be contracted. However, any such ${Q^x}_y$
can always be absorbed into the 3--hinge weights by redefining
${\widehat\Delta}^{xyz}\equiv{Q^{x}}_{x'}{Q^{y}}_{y'}
{Q^{z}}_{z'}\Delta^{x'y'z'}$; in the new frame ${{{\widehat
Q}^x}}_{{\phantom{x}}y}={\delta^x}_y$ reduces to our choice of a gluing
rule.}  For example, the total weight for the configuration of three
triangles shown in \fD\ is
$C_{abx}C_{cdy}C_{efz}\Delta^{xyz}$.

For the latter case, our gluing rule is provided by the operator
$S^x_{{\phantom{x}}y}$, whose
function is to change the direction of an arrow on a hinge.  For
example, the hinge shown in \tfig\fG\ is assigned the local weight
$S^x_{{\phantom{x}}x'}~\Delta^{x'yz}$. Because the two hinges
in\tfig\fH\ should have the same weight, we require the following
constraint between $S^x_{{\phantom{x}}y}$ and $\Delta^{xyz}$ to be
satisfied:
\eqn\CONDEL{
S^x_{{\phantom{y}}x'}\,\Delta^{x'yz} =
S^z_{{\phantom{z}}z'}\,S^y_{{\phantom{y}}y'}\,\Delta^{xz'y'}~. }
\ifig\fG{A configuration involving $S^x_{{\phantom{x}}x'}$.}{Fig-3-seven}
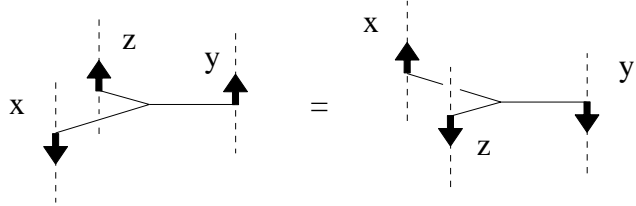
\ifig\fH{The constraint on $S^x_{{\phantom{x}}y}$ and
$\Delta^{xyz}$.}{Fig-3-eight}

\bigskip

The five weighting rules we have just described, $C_{xyz}$, $g^{xy}$,
$\Delta^{xyz}$, $h_{xy}$, and $S^x_{{\phantom{x}}y}$, are the
fundamental ingredients in our construction of LFTs. Provided that
they satisfy the consistency conditions \eCABCD, \eINDEP,
\eDBUBBLE, and \CONDEL, these data are sufficient to define a lattice
field theory; in terms of them, the partition function on an arbitrary
lattice $L$ is given by
\eqn\ePART{
Z(L)={\cal N}\, \prod_{f:\,{\rm faces}}\ \prod_{h:\,{\rm hinges}}
\prod_{e:\,<x,y>}
C_{a_1a_2\ldots a_k}(f)\, \Delta^{b_1b_2\ldots b_l}(h)~
\,{S^x}_y(e),}
where the third product is over pair of glued edges with arrows
reversed, all indices are assumed to be contracted, ${\cal N}$ is a
normalization factor, and $C_{a_1a_2\ldots a_k}(f)$ and
$\Delta^{b_1b_2\ldots b_l}(h)$ are defined in terms of triangular
decompositions of the $k$-gonal face and the $l$-hinge, as in eqs.\
\eNFACE\ and \eNHINGE.

\bigskip
Before  closing this subsection, we would like to make two remarks.

The first remark is that our construction of  lattice field theories
respects lattice duality. To understand this statement more
concretely, recall that the  dual lattice $\widetilde L$ of a given lattice
$L$ is composed of faces dual to the hinges of $L$ and hinges dual to the
faces of $L$.  Thus, given a LFT on the original lattice $L$ with data
($C_{xyz}$, $g^{xy}$, $\Delta^{xyz}$, $h_{xy}$, $S^x_{{\phantom{x}}y}$),
we can always define an equivalent lattice theory
on the dual lattice $\widetilde L$ with data (${\widetilde C}_{xyz}\equiv
\Delta^{xyz}$, ${\tilde g}^{xy}\equiv h_{xy}$, ${\widetilde \Delta}^{xyz}
\equiv C_{xyz}$, ${\tilde h}_{xy}\equiv g^{xy}$,
${\widetilde S}^x_{{\phantom{x}}y}\equiv {S^y}_x$).
It should be obvious that  the partition function of this
theory  on $\widetilde L$ is equal to the partition function of the
original theory on $L$.

Finally, we mentioned previously that we could have taken an alternative,
more general approach to defining lattice field theories on arbitrary
lattices.  By allowing the $C_{x_1x_2...x_k}$ (and the
$\Delta^{y_1y_2\ldots y_l}$) to be independent and not imposing
invariance under the subdivision of faces (hinges), we could have
defined a much broader class of lattice field theories, with an
infinitely larger set of weighting rules, at the expense of an
abundance of indices.  Our definition of an LFT imposes invariance under
face subdivisions at the outset, and thus restricts our discussion to a
class of lattice field theories
with a limited amount of topological invariance.  Alternatively,
subdivision invariance could have been imposed at a later stage, along
with invariance under other local lattice moves. The end result would
have been the same set of TLFTs that we will find later in this section.

\subsec{Algebra and Coalgebra Structures in 3D Lattice Field Theories}

We will now explain how the data $(C_{xyz}, g^{xy})$ and
$(\Delta^{xyz},h_{xy})$ can be used to define algebra and
coalgebra structures.\foot{See Appendix A for definitions of algebra,
coalgebra, and related concepts.} As in sec.\ 2, we introduce a vector space
$H\equiv {\oplus_{x\in X}}{\bf C} \phi_x$, with one basis element $\phi_x$
corresponding to each color $x\in X$. Given the data of the LFT
defined above, we can define a multiplication operation
on $H$ with  $C_{xy}^{{\phantom{xy}}z}\equiv
C_{xyz'}g^{z'z}$ as structure constants:
\eqn\eALG{\eqalign{
&m:~H\otimes H \rightarrow H \cr
&m\,(\phi_x\otimes\phi_y) \equiv \phi_x\cdot\phi_y\equiv
C_{xy}^{{\phantom{xy}}z} \phi_z~.\cr}}

The consistency requirements\ \eCABCD , imposed on any 2D face in our
definition of 3D LFTs, imply that the algebra thus defined is
associative and semisimple. As explained in subsec.\ 2.3, such an
algebra always contains a unit element $u$ satisfying
\eqn\eunitDEFA{\eqalign{
  u\,:&~{\bf C}\,\ni\,1\, \mapsto\, u(1)=u^x\phi_x\,\in\,H \cr
&u(1)\cdot\phi=\phi=\phi\cdot  u(1)\cr
i.e.~~~~&u^x\,{C_{xy}}^z={\delta_y}^z=u^x\,{C_{yx}}^z\cr}}
with $u_x\equiv C_{xa}^{{\phantom{xa}}a}$ and $u^x\equiv g^{xy}\,
u_y$. The last equation is equivalent to the 2D cone equation  
\etwocone, which follows directly from eq.\ \eCABCD.
For later convenience, we introduce in  \tfig\fM a
a graphical representation of $u_x$
as a 2D disc with a color index
$x$ on the boundary.  Eq.\ \eunitDEFA\ (or equivalently eq.\ \etwocone )
is thus depicted in \fM b, and is equivalent to the 2D cone move
pictured in \TWOunitone a and b.
\ifig\fM{Graphical representation of $u_x$ and \eunitDEFA.}{Fig-3-thirteen}

Similarly, we can use the hinge weights
${\Delta_x}^{yz}=h_{xx'}\Delta^{x'yz}$ to define a multiplication
operation on the vector space $\widetilde H\equiv {\oplus_{x\in X}}{\C}
\tilde\phi^x$:
\eqn\eDALG{\eqalign{
&\widetilde m :~~\widetilde H\otimes \widetilde H\rightarrow
\widetilde H \cr
&\widetilde m \, (\tilde\phi^y\otimes \tilde \phi^z) \equiv
{\Delta_x}^{yz} \tilde\phi^x~.\cr}}
It follows from the conditions for
invariance under decomposition into 3-hinges, eqs.\ \eINDEP\ and
\eDBUBBLE, that $\widetilde m$ is associative and semisimple, and
contains a unit element $\tilde u$,
\eqn\eq{
\tilde  u\,:~{\bf C}\,\ni\,1\, \mapsto\,
\tilde u (1)=\epsilon_x\tilde \phi^x\,\in\,\widetilde H }
with $\epsilon_x\equiv h_{xy}{\Delta_{u}}^{uy}$.

$\widetilde H$ is naturally identified with the dual vector space to
$H$, and through this identification, the multiplication
$\widetilde m$ naturally gives rise to a comultiplication
operation $\Delta$ on $H$, with the hinge data ${\Delta_x}^{yz}$ as
costructure constants:
\eqn\eCOALG{\eqalign{
&\Delta:~~H\rightarrow H\otimes H \cr
&\Delta(\phi_x) \equiv \lowdelta{yz}{x} \phi_y\otimes \phi_z~.\cr }}
The condition of coassociativity
\eqn\eq{
\left(\Delta\otimes 1_{H}\right)\circ\Delta(\phi_w)=\left(1_{H}
\otimes \Delta\right)\circ\Delta(\phi_w)}
follows from the associativity of $\widetilde m$ in $\widetilde H$.
Furthermore, duality directly
implies that the coalgebra \eCOALG\ is ``cosemisimple,'' and
contains a counit $\epsilon$ satisfying
\eqn\eunitDEFB{\eqalign{
\epsilon\,:&~H\,\ni\,\phi_x\,\mapsto\,\epsilon(\phi_x)=\epsilon_x\,\in\,
      {\bf C} \cr
&{\Delta_{x}}^{yz}\e_z={\delta_x}^y=\e_z{\Delta_{x}}^{zy}~,\cr}}
with $\epsilon_x\equiv\, h_{xy}\Delta_u^{{\phantom{u}}uy}$.
(See Theorem\ A.2 in Appendix A
for a more detailed discussion of the counit $\epsilon_x$.)
In\tfig\fN a, we introduce a
graphical representation for $\epsilon_x$ as a single edge with a loop
on the plane perpendicular to the edge,
and express eq.\ \eunitDEFB ,  or equivalently,
$\epsilon^z\,{\Delta_z}^{xy} =h^{xy}$, in \fN b.
\ifig\fN{Graphical representation of $\epsilon_x$ and
\eunitDEFB.}{Fig-3-thirteenb}

A brief remark may serve to clarify the relation between the algebra
and the coalgebra structures that we have defined in this subsection.
As we have explained, an equivalent lattice theory may be defined on the
dual lattice $\widetilde L$, whose weights are inherited from the original
lattice $L$.  In going to the dual lattice, hinges become faces and
faces become hinges, so the structure constants ${C_{xy}}^z$ and
${\Delta_{z}}^{xy}$ are simply interchanged, and the algebra and
coalgebra respectively determine the coalgebra and the algebra
of the dual theory on $\widetilde L$.

\subsec{Topological Invariance and the Correspondence to Hopf Algebras}

In the previous subsection, we showed how to use the data of a
three--dimensional
lattice field theory (LFT) to define algebra and coalgebra structures
on a vector space $H$.  There, the invariance of faces and hinges under
 subdivision implied that the algebra is associative
and semisimple, and the
coalgebra is coassociative and cosemisimple.
In this subsection, we show that upon requiring
our LFT to be  topologically invariant,
the algebra $(H; m, u)$ and the coalgebra $(H; \Delta,
\epsilon)$ are combined to form a constrained Hopf
algebra with antipode given by the arrow--reversing
operator $S$ appearing in
eq.\ \CONDEL .   (The definition and basic properties  of Hopf algebras
are discussed in Appendix A.)

To state the main result of this section more precisely, we first introduce an
operator $T$ with matrix elements
\eqn\eThreeA{  {T^x}_y \equiv h^{xz}\, g_{zy}~,}
and a numerical constant
\eqn\eThreeB{ \Lambda_0\equiv \epsilon_x\,{T^x}_y\,u^y =\epsilon^x\,u_x~.}
Then, we will prove the following theorem:
\medskip
\noindent {\bf Theorem 3.1.}\ {\it The set of 3D topological lattice field
theories (TLFTs) with data $(C_{xyz},\, g^{xy},\, \D^{xyz},\, h_{xy},\,
{S^x}_y)$ is in one-to-one correspondence with the set of Hopf algebras
$(H;\, m,\, u,\, \D,\, \epsilon,\,S)$ with antipode $S$ of the
following form:
\eqn\eThreeC{S={1\over{\Lambda_0}}\, T~.}}
\medskip
\noindent An important consequence of
eq.\ \eThreeC\ is that $S^2=1$
(see Theorem A.7).\foot{
This rather restrictive constraint should not be too surprising, since
the operation of reversing an arrow twice is trivial.
We will discuss the condition $S^2=1$ further in sec.\ 5.}

The proof of this theorem is divided into three parts.  In Step 1, we
will introduce two types of local moves -- the hinge move and the 3D
cone move -- and demand that the partition function $Z(L)$ be
invariant under each of them.  Invariance under the hinge move will be
shown to imply that the quintet $(H;\, m,\, u,\, \D,\, \epsilon)$
forms a bialgebra with a constraint to be given below in eq.\ (3.21).
On the other hand, invariance under the 3D cone move will guarantee
that the operator $S$ appearing in eq.\ \CONDEL\ satisfies the axiom
of antipode.  Thus, we can conclude that the data $(C_{xyz},\, g^{xy},\,
\D^{xyz},\, h_{xy},\,{S^x}_y)$ of  any LFT
invariant under the hinge and 3D cone moves defines a Hopf algebra $(H;\,
m,\, u,\, \D,\, \epsilon,\,S)$ with the constraint (3.21), which will
be shown in Theorem A.7 to be equivalent to the statement that the
antipode $S$ satisfies eq.\ \eThreeC .  In Step 2, we will prove that
these two local moves generate all local, topology--preserving
deformations of the lattice.  Thus, no additional constraints are
needed to guarantee that we have constructed a
topologically--invariant LFT, and we can conclude that we have a map
from the set of 3D TLFTs to the set of Hopf algebras with antipode
$S={1\over{\Lambda_0}}\, T$.  In Step 3, we will further show that
this map is bijective, {\it i.e.}, given a Hopf algebra with antipode
$S={1\over{\Lambda_0}}\, T$, we can always construct a 3D TLFT.  This
step will complete the proof of Theorem 3.1.

\medskip
\noindent ${\underline{\bf Step~1.}}$
\medskip

Our starting point is a simple identity, the ``hinge equation,'' which
expresses algebraically a necessary condition for topological
invariance.  It is derived from a topology--preserving move that takes
two triangles glued to each other along two hinges, and collapses them
homotopically to a single triangle plus a hinge (and vice versa).  Picture
the triangles as forming a conical tube, as in\tfig\fO , possibly with
additional faces attached to the two hinges.  Then provided that none
of these additional faces
 are attached to the hinges in the {\it
interior} of the tube, we can squeeze it from the bottom
and flatten it as we go up,
like a tube of toothpaste, pushing out any structure
attached to the two open edges (labeled $a$ and $b$ in
{\fO}) and ending up with a single flat
triangle attached to a hinge.  We call this operation the
``hinge move,'' and call the equation expressing the invariance under
the hinge move the hinge equation. (A similar move, which collapses
two triangles glued together along all three edges to a single
triangle, is derived from the hinge move in Appendix B.)  The hinge
equation  is reminiscent of the two--dimensional bubble equation
\twobub, which told us how to collapse two triangles sharing two
edges to a single edge. (See \TwoB.)

The expression of invariance under the hinge move
in terms of the local weights can be
deduced from a careful inspection of \fO\ as follows:
\eqn\hingeid{
\lowdelta{pq}{x}~\lowdelta{rs}{y}~C_{apr}~C_{bqs} = \Lambda\,
C_{xyz}\,{\Delta^z}_{ba}~.}
Here, a numerical factor $\Lambda$ is introduced to count the number of
3D cells,  and will be determined below.
\ifig\fO{The hinge move.}{Fig-3-fourteen}

To obtain an algebraic interpretation of eq.\ \hingeid , we first need
the following identities, whose proofs are contained in Appendix B:
\eqn\eLM{\eqalign{
&u^{x}\,\Delta_x^{{\phantom{x}}yz}=u^y\,u^z\cr
&C_{xy}^{{\phantom{xy}}z} \epsilon_z = \epsilon_x\,\epsilon_y~.\cr}}
\medskip
\noindent Multiplying both sides of eq.\ \hingeid\ by $u^x$ and using
\eLM, we find
\eqn\hingeidGene{
T^a_{{\phantom{a}}r}\,T^b_{{\phantom{b}}s}\,\Delta_y^{{\phantom{y}}rs}
=\Lambda\, T^z_{{\phantom{z}}y}\,\Delta_{z}^{{\phantom{z}}ba}~,}
where we have used $T^a_{{\phantom{a}}r}= h^{aa'}g_{a'r}$.
Note that eq.\ \hingeidGene\ is equivalent to the
statement that ${1\over{\Lambda}}\,T$ is
a coalgebra antimorphism \Abe.  Next we substitute eq.\ \hingeidGene\ into
eq.\ \hingeid , and obtain
\eqn\conditionII{
\Delta_x^{{\phantom{x}}pq}\,\Delta_y^{{\phantom{y}}rs}\,
C_{pr}^{{\phantom{pr}}a}\,C_{qs}^{{\phantom{qs}}b}=
C_{xy}^{{\phantom{xy}}z}\,\Delta_z^{{\phantom{z}}ab}.}
This equation is equivalent to the statement that
comultiplication $\Delta$ is an algebra
morphism with respect to multiplication $m$, $\Delta\,(\phi_x\cdot\phi_y)
=\Delta(\phi_x)\cdot\Delta(\phi_y)$.  More precisely,
\eqn\eq{
\Delta\circ m~(\phi_x\otimes\phi_y)=(m\otimes m)\circ (1\otimes\tau\otimes 1)
\circ (\Delta\otimes\Delta)~(\phi_x\otimes\phi_y)~,}
where $\tau$ is  the twist mapping:
\eqn\eq{
\tau\,:~H\otimes H\,\ni\,\phi_a\otimes\phi_b\,\mapsto\,\phi_b\otimes\phi_a\,
\in\, H\otimes H~.}
Furthermore, multiplying eq.\ \conditionII\ by $u^y\,\epsilon_b$, and using
eqs.\  \eunitDEFA, \eunitDEFB\ and \eLM , we obtain
${\delta_x}^a = {\delta_x}^a\, u^y\,\epsilon_y$, or
\eqn\eeUNIT{
 \epsilon_x\, u^x =1~. }
Remarkably,  eqs.\ \eLM , \conditionII\ and \eeUNIT\ are exactly
the conditions for $(H;m,u,\Delta,\epsilon)$
to be a bialgebra (see Appendix A).  That is,
\medskip
\noindent{\bf Lemma 3.2.}\
{\it The hinge equation  \hingeid\ and
eq.\ \eLM\ imply that the quintet
$(H;m,u,\Delta,\epsilon)$ forms a bialgebra with the constraint \hingeidGene .}
\medskip

Next, we will show that the lattice operator $S$ defined in subsec.\ 3.1
is the antipode of a Hopf algebra $(H;m,u,\Delta,\epsilon, S)$,
by considering a configuration
in which  two edges of a single triangle are glued together on a
hinge, and imposing invariance under a particular local move that removes
such a triangle. Because a triangle of this sort looks like a cone, we
will refer to this move as the ``3D cone move.'' Its effect is to
contract the surface of the cone down to a minimal--area surface
bounded by the edges of the original triangle; the result looks like a
vertical edge, labeled $x$ in \tfig\ADDfig , attached at the bottom to
a disc with boundary $y$.
\ifig\ADDfig{3D cone move.}{Fig-3-conemove}

Invariance under the particular cone move shown in \ADDfig\
implies that
\eqn\eCONEIDA{
\Delta_{x}^{{\phantom{x}}ab}\,S^c_{{\phantom{c}}b}\,
C_{ac}^{{\phantom{ac}}y} = u^y \,\eps_x ~.   }
By changing the orientations of the edges $x$ and $y$, and imposing
invariance under three nearly identical cone moves,
we obtain the following set of identities:
{\phantom{ {\eqnn\eCONEIDB}{\eqnn\eCONEIDC}{\eqnn\eCONEIDD} }}
$$\eqalignno{
&\Delta_{x}^{{\phantom{x}}ab}\,S^c_{{\phantom{c}}a}\,
C_{cb}^{{\phantom{bc}}y} = u^y \,\eps_x {\phantom{~.}}&\eCONEIDB \cr
&\Delta_{x}^{{\phantom{x}}ab}\,S^c_{{\phantom{c}}b}\,
C_{ca}^{{\phantom{ac}}y} = u^y \,\eps_x {\phantom{~.}}&\eCONEIDC \cr
&\Delta_{x}^{{\phantom{x}}ab}\,S^c_{{\phantom{c}}a}\,
C_{bc}^{{\phantom{bc}}y} = u^y \,\eps_x~.&\eCONEIDD \cr  }$$
Eqs.\ \eCONEIDA\ and \eCONEIDB\ are  the defining axioms for
antipode $S$ of a Hopf algebra, and \eCONEIDC\ and \eCONEIDD\ are
equivalent to the property $S^2=1$ (see Appendix A):
\eqn\eANTIPODE{\eqalign{
m\circ(S\otimes 1)\circ \Delta
 &= m\circ(1\otimes S)\circ \Delta =  u\circ \epsilon\cr
&S^2=1~.\cr}}

Furthermore, due to Theorem A.7, the constraint \hingeidGene\ on $T$
implies that $\Lambda=\Lambda_0$, and
 the antipode $S$ has the following form:
\eqn\eSL{S={1\over{\Lambda_0}}\,T~,}
  from which we can again conclude that $S^2=1$.  Thus, we have proven
\medskip
\noindent{\bf Lemma 3.3.}\ {\it
Invariance under the hinge move, 3D cone moves, and Eq.\ \eLM\ imply
that the sextet $(H;m,u,\Delta,\epsilon, S)$ forms a Hopf algebra with
 antipode $S={1\over{\Lambda_0}}\,T$.}
\medskip

\noindent Note that the consistency condition \CONDEL\ is automatically
satisfied since
\eqn\CONS{\eqalign{
S^y_{~b}\,S^z_{~c}\,\Delta^{xcb}&=S^y_{~b}S^z_{~c}\,h^{xx'}\,
\Delta_{x'}^{{\phantom{x'}}cb}
=h^{xx'}\,S^{a'}_{{\phantom{a'}}x'}\,\Delta_{a'}^{{\phantom{a'}}yz}\cr
&=h^{xx'}\,S^{a'}_{{\phantom{a'}}x'}\,h_{a'a}\,\Delta^{ayz}
={1\over{\Lambda_0}}\,h^{xx'}\,h^{a'b}\,g_{bx'}\,h_{a'a}\,
\Delta^{ayz}\cr
&=S^x_{~a}\,\Delta^{ayz}~.\cr}}
by eqs.\ \eThreeA, \hingeidGene, and \eSL.
\medskip
\noindent ${\underline{\bf Step~2.}}$
\medskip

The hinge and 3D cone moves are simple yet powerful moves relating
topologically equivalent lattices. Another move we have discussed is
the subdivision of a polygonal face into two or more polygons.  We
claim that these moves are sufficient to generate all lattices in a
given topological equivalence class. The method of our proof will be
to show that these moves generate a set of standard moves, which are
sufficient by Alexander's theorem.

\medskip
\noindent{\bf Theorem 3.4.}\ {\it Any local deformation of a 3D lattice can be
generated by a sequence of  hinge, 3D cone, and face--subdivision moves.}
{\phantom{ {\tfig\ADDTHREEone}  {\tfig\ADDTHREEtwo}  }}
\smallskip
\item{}{\noindent{\it{Proof.}}~~
Given two topologically equivalent lattices $L$ and $L'$, we must show
the existence of an interpolating sequence of topology--preserving
deformations.  We begin by triangulating all polygonal faces in $L$
and $L'$, and using hinge and 3D cone moves to eliminate any pairs of
triangles with two edges in common as well as any single triangles
with two edges meeting at a hinge. The resulting ``good''
triangulations $\bar L$ and ${\bar L}'$ have the property that at
least three triangles will always meet at any vertex of any
polyhedron.  We would then like to show that hinge moves and
face--subdivision moves are sufficient to decompose each resulting
polyhedron into tetrahedra.  Let us postpone the proof of this
statement for the moment, and see how the theorem follows once a
tetrahedral decomposition has been constructed.
\ifig\ADDTHREEone{(2,3) move.}{Fig-3-23one}
Then $\bar L$ and
${\bar L}'$ can be respectively decomposed into two topologically
equivalent tetrahedral lattices $\Gamma$ and $\Gamma'$.  At this
point, we invoke a version \Mat\Boulatov\ of Alexander's theorem
\ALEX, according to which any two topologically equivalent tetrahedral
lattices are related by a sequence of special local lattice moves,
known as ``(2,3)'' and ``bubble'' moves (\ADDTHREEone\ and
\ADDTHREEtwo ).  The (2,3) move relates a configuration
containing two tetrahedra joined along a face to a configuration in
which the two tetrahedra have been replaced by three tetrahedra joined
along a hinge (which is dual to the original face), while the bubble
move collapses two tetrahedra sharing three faces to a single
triangle. Both types of moves are explained in detail in Appendix B,
and shown to be generated by sequences of hinge and face--subdivision
moves. Thus, the
sequence of (2,3) and bubble moves interpolating between $\Gamma$ and
$\Gamma'$ can be reduced to a sequence of hinge moves, which can in
turn be incorporated into a sequence of hinge, 3D cone, and face--subdivision
moves interpolating between $L$ and $L'$.}
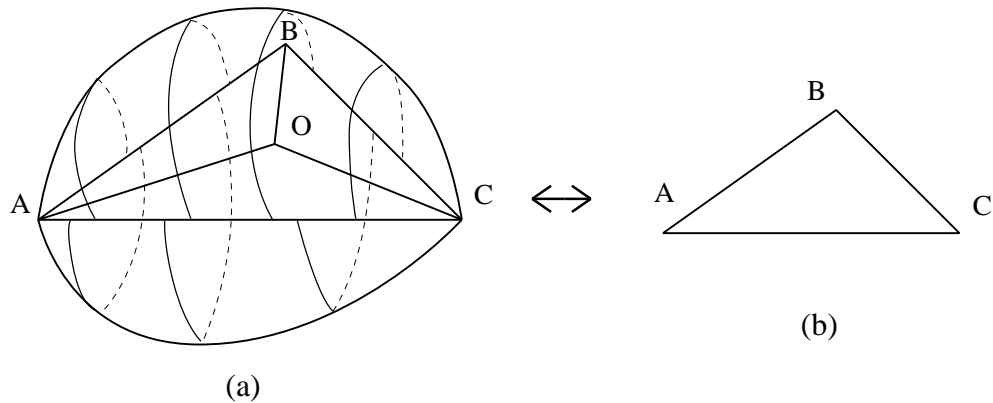
\ifig\ADDTHREEtwo{3D Bubble move}{Fig-3-bubblemove}

\item{}{\indent It remains to prove our assertion that hinge and
face--subdivision moves alone are
sufficient to decompose a triangulated polyhedron into tetrahedra.  We
will perform an induction on the number of faces $F$ of the polyhedron
$P$.  Recall that our ``good'' triangulation has been chosen so that
at least three triangles always meet at any vertex of $P$. Let us choose
a vertex $O$ at which $m$ triangles meet (\tfig\LEMone\ ).}
\ifig\LEMone{$m$ triangles meeting at a vertex $O$.}{Fig-3-lem4one}
\item{}{\indent In order to visualize three--dimensional lattice
operations more clearly, we raise the position of the vertex
$O$ as shown in\tfig\LEMthree a, relative to \LEMone . Now
consider the two adjacent triangles $OA_1A_2$ and $OA_2A_3$, singled
out in \tfig\LEMtwo. By applying the hinge move to these triangles
twice, we obtain two additional internal triangles with the same
vertices, joined by a face--gluing operator. (This process of
``inflation'' is described in detail in Lemma B.4.)
After acting on these internal triangles with a (2,2) move,
the resulting internal triangles $OA_1A_3$ and $A_1A_2A_3$
together with the two external triangles
$OA_1A_2$ and $OA_2A_3$ make up the four faces of a tetrahedron.
We now repeat this process of tetrahedron--formation on the
successive pairs of adjacent triangles $OA_1A_{k}$ and
$OA_{k}A_{k+1}$ ($k=3,\,\ldots\, ,m-1$),
obtaining finally $m-2$ interior tetrahedra shown in
{\LEMthree}b, whose external faces are the original $m$ triangles
that we started with.\foot{More precisely, the resulting configuration
includes an inflated triangular tube $OA_mA_1$, which can be flattened
by a hinge move, as described in Lemma B.2.} After removing
these tetrahedra in \LEMthree c, we are left with a polyhedron
with $F-2$ faces (also obeying the ``good'' condition), to which the
induction hypothesis can be applied.}
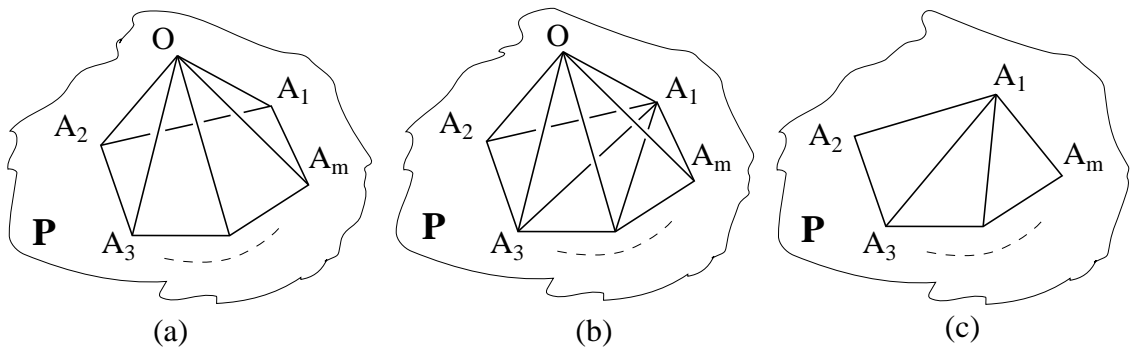
\ifig\LEMthree{Inductive construction of a tetrahedral
decomposition.}{Fig-3-lem4three}
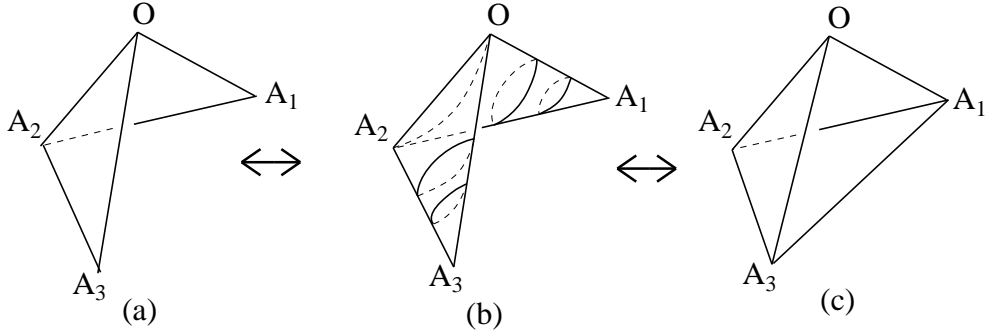
\ifig\LEMtwo{Construction of a tetrahedron using hinge
moves.}{Fig-3-lem4two}
\item{}{\indent We continue reducing the number of faces until
we eventually reach a good polyhedron with four faces, that is, a
tetrahedron.  (Here we make use of the fact that our polyhedral
3--cells are all contractible.) At this point, the construction of a
tetrahedral decomposition of $P$ is complete, and so is the proof.
\ \  (QED)}

\bigskip

\noindent ${\underline{\bf Step~3.}}$
\bigskip

The preceding lemma implies the existence of a map from the set of 3D
TLFTs to the set of Hopf algebras with antipode
$S={1\over{\Lambda_0}}\,T$.  In this step, we further prove that this
map is bijective.  Given a Hopf algebra $(H;m,u,\Delta,\epsilon, S)$
with antipode $S={1\over{\Lambda_0}}\,T$, we first define the metric
and the cometric by
\eqn\eq{\eqalign{
g_{xy}&=C_{xa}^{{\phantom{xa}}b}\, C_{yb}^{{\phantom{yb}}a}\cr
h^{xy}&=\Delta_a^{{\phantom{a}}bx}\,\Delta_b^{{\phantom{b}}ay}~,\cr}}
which can be used to define the cyclic local weights $C_{xyz}\equiv
g_{zz'}\,C_{xy}^{{\phantom{xy}}z'}$ and $\Delta^{xyz}\equiv
h^{xx'}\Delta_{x'}^{{\phantom{x'}}yz}$.  Since $S$ is a coalgebra
antimorphism, we can show that $S^q_{~r}$ satisfies eq.\ \CONDEL,
using the same sequence of manipulations that appeared in eq.\ \CONS .
In other words, we can consistently assign a local weight to a hinge
whose arrows are not in the same direction.  A 3D LFT can thus be
constructed from the data ($C_{xyz},\, g^{xy},\,h_{xy},\,
\Delta^{xyz},\, S^q_{~r})$.  Moreover, the Hopf algebra axioms
and the form of the antipode will ensure that this 3D LFT satisfies
the hinge equation  \hingeid\ and the cone identities \eCONEIDA
--\eCONEIDD .  Thus, given a Hopf algebra with
$S={1\over{\Lambda_0}}\,T$, we can always construct a 3D TLFT.

\newsec{\bf Example: $H$=$\C [G]$ and the Ponzano--Regge Models}
\global\figno=1

We now apply the results of the preceding  section to the
group ring $H={\C}[G]$ of a compact group:
\eqn\eq{
H={\C}[G]={\bigoplus_{x\in G}}~ {\C}~\phi_x~.}
Although $\C [G]$ is one of the simplest examples of a Hopf algebra,
it contains enough structure to provide a good illustration of the
concepts of Section 3, and to lead to an interesting topological
lattice field theory.

In subsec.\ 4.1, we review the Hopf algebra structure of $\C [G]$. Next, in
subsec.\ 4.2, we  construct the TLFT associated
with $H=\C [G]$ and relate it to a
lattice gauge theory with heat--kernel action. Finally,
in subsec.\ 4.3, we will further relate this TLFT to the lattice theory
of Ponzano and Regge \refs{\RP,\Boulatov}.

\subsec{$H={\C}[G]$ as a Hopf algebra}

Let $G$ be a compact group.
$H={\C}[G]$ becomes a Hopf
algebra if we define (bi-)linear maps $m,~u,~\Delta,~\varepsilon$
and $S$ as follows:

\indent\indent (i) multiplication
\eqn\gone{m:~H\otimes H
\owns \phi_x\otimes\phi_y~\mapsto~\phi_{xy} \in H}
\indent\indent (ii) unit
\eqn\gtwo{ u:~{\C}\owns 1~\mapsto~\phi_{1_G} \in H}
\indent\indent  (iii) comultiplication
\eqn\gthree{\Delta:~H \owns
\phi_x ~\mapsto ~\phi_x\otimes\phi_x\in H\otimes H}
\indent\indent  (iv) counit
\eqn\gfour{\varepsilon:~H\owns\phi_x~\mapsto~1\in {\C}}
\indent\indent  (v) antipode
\eqn\gfive{S:~H\owns\phi_x~\mapsto~\phi_{x^{-1}}\in {H}~.}
It is easy to show that these definitions satisfy the Hopf algebra
axioms, which are given in Appendix A.  This Hopf algebra is
cocommutative, and has an antipode $S$ obeying $S^2=1$.

Eqs.\ \gone\ - \gfour\ determine the structure constants $C_{xy}^{~~z}$
and the costructure constants $\Delta_x^{~yz}$ as
\eqn\defin{\eqalign{
&C_{xy}^{~~z} = \delta(xy,z)\cr
&{\Delta_x}^{yz}= \delta(x,y)~\delta(x,z)~.\cr}}
The metric and cometric are obtained by combining eq.\ \defin\ with
eqs.\ \eCABCD, \eDBUBBLE , \eunitDEFA\ and \eunitDEFB :
\eqn\eq{\eqalign{&g_{xy}=\delta(x,y^{-1})\cr
&u^x=\delta(x,1_G)\cr
&h^{xy}=\gn\delta(x,y)\cr
&\epsilon_x=1\cr}
   ~~~~~~~~~\eqalign{&g^{xy}=\delta(x,y^{-1})\cr
&u_x=\delta(x,1_G)\cr
&h_{xy}={1\over{\gn}} \delta(x,y)\cr
&\epsilon^x=|G|\cr}}
with $\gn\equiv {\rm card}(G)=\delta(x,x)$.\foot{Since $\gn$ is infinite in
a continuous group, the following discussion is of a somewhat formal nature.}
The numerical constant $\Lambda_0$ appearing in eq.\ \eThreeB\ is determined as
\eqn\eq{\Lambda_0=\int dx~\epsilon^x u_x=|G|~.}

It is easy to see that the condition \eThreeC\ is
satisfied for this Hopf algebra:
\eqn\eq{
{1\over{\Lambda_0}}\, {T^x}_y= {1\over{\Lambda_0}} \int {\rm d}z\,h^{xz}\,
g_{zy}=\d (x,y^{-1}) = {S^x}_y~.}
Thus, due to Theorem 3.1, we can construct a TLFT with the
following data $(C_{xyz},\,g^{xy},\,\D^{xyz},\,h_{xy},\, {S^x}_y)$:
\eqn\eq{\eqalign{
C_{xyz}&\equiv \int {\rm d}z'~{C_{xy}}^{z'}g_{z'z}=\delta(xyz, {1_G})\cr
g^{xy}&=\delta(x,y^{-1})\cr
\Delta^{xyz}&\equiv \int {\rm d}x'~h^{xx'}{\Delta_{x'}}^{yz}=
|G|\,\delta(x,y)\delta(x,z).\cr
h_{xy}&={1\over{\gn}}\, \d(x,y)\cr
{S^x}_y&=\d (x,y^{-1})~.\cr}}

The contribution from an arbitrary polygonal face
can be evaluated as in eq.\ \eNFACE:
\eqn\eCHIGHER{\eqalign{
C_{x_1x_2\ldots x_n}
&=\int {\rm d}a_1{\rm d}a_2...{\rm d}a_{n}{\rm d}a'_1
{\rm d}a'_2...{\rm d}a'_n~g^{a_1a'_1}...g^{a_na'_n}\cr
&\qquad\qquad\times C_{a_1x_1a_2}C_{a'_2x_2a_3}\,\cdots\,
C_{a'_nx_na'_1}\cr
&=\delta(x_1x_2...x_n,1_G)~.\cr } }
Hence, in configurations with nonzero weight, the product of the link
variables around each two-dimensional cell is constrained to equal the
unit element of $G$.

On the other hand, the contribution of a hinge joining $n$ faces,
as given by the generalized hinge operator of eq.\ \eINDEP,
is easily seen to be
\eqn\eDHIGHER{
\Delta^{x_1x_2...x_n} =
  |G|\,\delta(x_1,x_2)\,\delta(x_1,x_3)\, ...\,\delta(x_1,x_n)~.}
Eq.\ \eDHIGHER\ implies that the edges which are connected to the same
hinge have the same value in $G$.

Using  these expressions for $C_{x_1x_2\ldots x_n}$
and $\Delta^{x_1x_2...x_n}$ we can now write down the
partition function $Z(L)$ for a given cellular decomposition $L$
as\foot{
The normalization factor ${\cal N}$ can be calculated as
${\gn}^{-N_3(L)}$ from the factors of $\Lambda=\gn$ appearing in
the (2,3) and bubble moves, as described in Appendix B.}
\eqn\ddeone{
Z(L)={1\over{{\gn}^{N_3(L)}}} \int \prod_{l\in C_1}{\rm d}x_l
\prod_{f\in C_2} \delta\left(\prod_{l\in f}x_l~,1_G\right)~.}
Here, $C_r$ denotes the set of $r$-dimensional cells, and $N_r(L)$ is
the number of $r$-dimensional cells; $N_r(L)\equiv\vert C_r \vert$.
As was explained in {sec.\ 3}, the direction of each link variable
 $x_l~(l\in C_1)$ should be fixed before its integration, although the
final answer is independent of the direction.

\subsec{Relation to topological lattice gauge theory}

The framework and results of the previous subsection are strongly
reminiscent of lattice gauge theory (LGT), where one also associates a
group element to each link of the lattice.  In fact, the TLFT
 based on $H={\C}[G]$ reproduces the zero coupling limit of LGT \Boulatov\
in the following way.

First, we define the partition function of the heat--kernel LGT action
for a cellular decomposition $L$, at inverse temperature $\beta$
(related to the bare coupling $g_0$ as $\beta\sim 1/g_0^2)$, to be \CREUTZ
\eqn\deone{\eqalign{
Z_{\beta}^{\rm LGT}(L)
&\propto\int \prod_{l\in C_1} {\rm d}x_l \prod_{f\in C_2}
e^{-S_{\beta}(U_f)} \cr
e^{-S_{\beta}(U_f)}&\equiv \sum_j~ d_j~ \chi_j(U_f)~
e^{-Q_j/\beta}.\cr}}
Here, $U_f$ is a plaquette variable for a face $f\in C_2$ $\left(
U_f\equiv \prod_{l\in f} x_l\right)$, and $d_j,~\chi_j$ and $Q_j$ are
the dimension, the character and the quadratic Casimir of an
irreducible representation $j$, respectively. Now taking the zero
coupling limit $\beta\rightarrow \infty$ ($g_0\to 0$)
in eq.\ \deone , we find
\def\firstl#1\over#2{\mathrel{\mathop{\null#2}\limits^{\scriptstyle#1}}}
\eqn\eq{
e^{-S_{\beta}(U_f)}
  ~{\firstl{\beta\to\infty}\over{\longrightarrow}}~
\sum_j~d_j \chi_j(U_f)
=\delta(U_f,1_G)=\delta\left(\,\prod_{l\in f}x_l, 1_G\right)~,}
and thus ${\lim_{\beta\rightarrow\infty}} Z_{\beta}^{\rm LGT}(L)$
actually yields the same expression as eq.\ \ddeone\ up to an irrelevant
normalization factor.

The topological nature of the zero coupling limit of LGT has a simple
explanation in the continuum limit. As $g_0 \to 0$,
the kinetic term
\eqn\eKIN{
{1\over{g_0^2}}\int d^3x~ {\rm Tr~}F_{\mu\nu}F^{\mu\nu}}
becomes peaked about flat connections with $F_{\mu\nu}\equiv 0$, for
which the action vanishes. The path integral then collapses to an
integral over gauge--inequivalent flat connections. Likewise, in the
TLFT for $\C [G]$, the form of the generalized structure constants
$C_{x_1\ldots x_k}=\delta(x_1x_2...x_k,1_G)$ imposes a lattice analogue of the
flatness constraint on those configurations that contribute with
nonzero weight. Thus, in the continuum limit, the partition function
\ddeone\ also becomes an integral over flat connections.

\subsec{ Relation to the Ponzano--Regge Model}

In this subsection, we show that
the partition function of the TLFT associated with $H={\C}[G]$ for
$G={\rm SU}(2)$ reproduces that of Ponzano and Regge \RP , which is equal to
the classical ($q$=1) Turaev--Viro invariant\ \TV.
Our derivation follows the logic of ref.\ \Boulatov.

The Ponzano--Regge model is defined as follows.  We first take a tetrahedral
cell decomposition $L$ and color the links with representations of SU(2).
To each tetrahedron, we then assign a numerical weight, the 6$j$-symbol of
SU$(2)$ corresponding to the representations on the
six edges of the tetrahedron. The partition function is then
(essentially) the sum over colorings of the product of the tetrahedral
weights,
\eqn\eTV{
Z_{\rm PR}({L}) \equiv {1\over{{\gn}^{N_0({L})}}}\sum_{\{ j \} }
\left(\prod_{{ l}\in { C_1}} d_{j_{ l}}\right)
\prod_{{ T}\in { C_3}}{\Bigl\lbrace
\matrix{j_{ T_1}&j_{ T_2}&j_{ T_3}\cr
j_{ T_4}&j_{ T_5}&j_{ T_6}\cr} \Bigr\rbrace}~.}
(The notation of this expression will be explained below.)

We will show that the partition function \eTV\ of the tetrahedral
complex $L$ is equal to the partition function of the TLFT
defined on the same lattice for the dual Hopf algebra $\widetilde H=
\widetilde{\C [G] }$ (see the remark after Corollary A.6).   The dual
Hopf algebra $\widetilde H=\widetilde{\C[G]}$ has the following
structure constants:
\eqn\eSC{\eqalign{
{\widetilde C}_{xy}^{{\phantom{xy}}z}&=\d(x,z)\, \d(y,z)\cr
{\widetilde\Delta}_x^{{\phantom{x}}yz}&=\d(x,yz)\cr}}
as is easily read out from eq.\ \defin. The lattice theory is thus
defined by the following data:
\eqn\eDATA{\eqalign{
{\widetilde C}_{xyz}&=|G|\, \d(x,y)\, \d(x,z)\cr
{\tilde g}^{xy}&={1\over |G|}\, \d(x,y)\cr
\tilde u^x&=1\cr}~~~~
\eqalign{{\widetilde \Delta}^{xyz}&=\d(xyz,1_G)\cr
{\widetilde h}_{xy}&=\d(xy,1_G)\cr
\tilde \epsilon_x&=\d(x,1_G)~.\cr}    }
For the generalized weights we have
\eqn\eGW{\eqalign{
\widetilde C_{x_1x_2\ldots x_n}&=|G|\, \d(x_1,x_2)\,\cdots\,\d(x_1,x_n)\cr
\widetilde \Delta^{x_1x_2\ldots x_n}&=\d(x_1x_2\cdots x_n,1_G)~.\cr}}
Due to the delta functions, the only configurations that contribute
with nonzero weight are those in which every link around every face is
mapped to the same group element, and in which the product of the
group elements on all the edges joined at any hinge is equal to  $1_G$.
Therefore, the partition function of this TLFT is equal to\foot{The
normalization factor  ${|G|}^{-N_0(L)}$ can be obtained from the
normalization factor in eq.\ \ddeone\
by a duality transformation.}
\eqn\ddDUALP{
Z(L)={1\over{{\gn}^{N_0(L)}}} \int \prod_{f\in C_2}{\rm d}x_f
\prod_{h\in C_1} \delta\left(\prod_{f~around~h}x_{f}~,1_G\right)~,}
where $x_f$ are   the face variables around the hinge $h$.

To evaluate this partition function \ddDUALP ,
we first rewrite the delta function  on the hinge $h$
in terms of the matrix elements of $x\in G$ in
an irreducible representation $j_h$:
\eqn\eq{
D^{j_h}_{mn}(x)\equiv \langle j_h,m\vert x\vert j_h,n\rangle~,}
where $m$ and $n$ are weight vectors in the representation $j_h$.
Recall here that they satisfy the following equations\ \MESS :
\eqn\formula{\eqalign{
\int {\rm d}x~D^{j}_{mn}(x){D}^{j'}_{m'n'}(x^{-1})
&={1\over{d_{j}}}~\d^{jj'}\d_{mm'}\d_{nn'}\cr
\sum_mD^{j}_{lm}(x){D}^{j}_{mn}(y)&=D_{ln}^j(xy)\cr
\sum_mD^j_{mm}(x)&=\chi_j(x)~.\cr}}
Thus, by using eq.\ \formula\ we can re-express the $\delta$-function in
eq.\ \ddDUALP\ as a product of matrix elements:
\eqn\fourtwo{\eqalign{
\delta\left(\prod_{{f}~{\rm around}~{h}}x_{f}~,
1_G\right)&=\sum_{j_{h}}~d_{j_{h}}~\chi_{j_{h}}
(x_1x_2...x_n)\cr
&=\sum_{j_{h},\{ m_i\}}
d_{j_{h}}~D_{m_1m_2}^{j_{h}}(x_1)
D_{m_2m_3}^{j_{h}}(x_2)...D_{m_km_1}^{j_{h}}(x_n)~.\cr}}
Substituting this expression back to eq.\ \ddDUALP , the partition
function becomes
\eqn\ePF{
Z(L)={1\over{{\gn}^{N_0(L)}}}
\int \prod_{f\in {C}_2}{\rm d}x_{f}
\prod_{h\in {C}_1} ~\sum_{j_{h},\{ m_i\}}
d_{j_{h}}~D_{m_1m_2}^{j_{ h}}(x_1)
D_{m_2m_3}^{j_{ h}}(x_2)...D_{m_km_1}^{j_{ h}}(x_n)~.}
Now since all faces are triangular, each $x_{f}$ appears three
times in the product over hinges. So we can rewrite the product as a
product over faces:
\eqn\fouragain{\eqalign{
Z(L) ={1\over{{\gn}^{N_0({L})}}}&\sum_{\{  j \} }
\left(\prod_{{h}\in { C_1}} d_{j_{h}}\right)\cr
&\sum_{\{ m_{{ f}_i},n_{{ f}_i}\} }
\int \prod_{{ f}\in { C_2}}
{\rm d}x_{ f}~
D_{m_{ f_1}n_{ f_1}}^{j_{ f_1}}(x_{ f})
D_{m_{ f_2}n_{ f_2}}^{j_{ f_2}}(x_{ f})
D_{m_{ f_3}n_{ f_3}}^{j_{ f_3}}(x_{ f})~,\cr}}
where the indices $\{ m_{{ f}_i},n_{{ f}_i}\}$ (for $i=1,2,3$)
should be contracted in such a way that they form a trace around each
hinge $h$ as is shown in eq.\ \fourtwo . In eq.\ \fouragain, we notice
that for each face variable, $x_{f}$, there are three representations
$j_{f_i}$, for $i=1,2,3$, associated with the three edges.

In the following, we restrict ourselves to the case $G=SU(2)$;
{$m=-j,~-j+1,...,~j-1,~j$} and $d_j=2j+1$.  In this case, we have
the additional equations\ \MESS
\eqn\use{
\int {\rm d}x~D^{j_1}_{m_1n_1}(x)
	D^{j_2}_{m_2n_2}(x) D^{j_3}_{m_3n_3}(x)
={\left( \matrix{j_1&j_2&j_3\cr
m_1&m_2&m_3\cr}\right)}  {\left( \matrix{j_1&j_2&j_3\cr
	n_1&n_2&n_3\cr}\right)}}
and
\eqn\gparity{
D^j_{mn}(x^{-1})=(-1)^{(j+m)+(j+n)} D_{-n-m}^j(x)~.}
Here, ${\left( \matrix{j_1&j_2&j_3\cr m_1&m_2&m_3\cr}\right)}$ is
Wigner's $3j$-symbol
\eqn\eq{
{\left( \matrix{j_1&j_2&j_3\cr
                 m_1&m_2&m_3\cr}\right)}=
{(-1)^{j_1-j_2-m_3}\over{{\sqrt{d_j}}}}
\langle j_1j_2,~m_1m_2\vert j_3,-m_3\rangle~,}
and eq.\ \gparity\ expresses the $G$-parity invariance of $SU(2)$.
Thus by integrating over $x_f$, we obtain two $3j$-symbols for each face $f$.
The product over faces of $3j$-symbols can be
rearranged into a product over tetrahedra of 6$j$-symbols
\eqn\eSIXJ
{\eqalign{
&\left\{ \matrix{j_{1}&j_{2}&j_{3}\cr
j_{4}&j_{5}&j_{6}\cr} \right\}
\equiv\sum_{\{m_i\}}(-1)^{j_4+j_5+j_6+m_4+m_5+m_6}\cr
&\cdot
{\left( \matrix{j_1&j_2&j_3\cr  m_1&m_2&m_3\cr}\right)}
{\left( \matrix{j_5&j_6&j_1\cr  m_5&-m_6&m_1\cr}\right)}
{\left( \matrix{j_6&j_4&j_2\cr  m_6&-m_4&m_2\cr}\right)}
{\left( \matrix{j_4&j_5&j_3\cr  m_4&-m_5&m_3\cr}\right)}~.\cr}}
The four faces whose $3j$-symbols appear here form a tetrahedron, as shown
in\tfig\FOURC .
\ifig\FOURC{A tetrahedron with edge variables $j_i$.}{Fig-4-three}
Performing the integration in eq.\ \fouragain\
and using eqs.\ \use\ and \gparity , we finally obtain the partition function
on $L$ \Boulatov
\eqn\fouragain{
Z(L)= {1\over{{\gn}^{N_0({L})}}}\sum_{\{ j \} }
\left(\prod_{{h}\in { C_1}} d_{j_{h}}\right)
\prod_{{ T}\in { C_3}}{\Bigl\lbrace
\matrix{j_{ T_1}&j_{ T_2}&j_{ T_3}\cr
j_{ T_4}&j_{ T_5}&j_{ T_6}\cr} \Bigr\rbrace}~.}
\noindent The right hand side of eq.\ \fouragain\ is nothing but the
partition function of the Ponzano--Regge model \eTV .  Furthermore,
since the Ponzano--Regge model is equivalent to the classical limit
($k\rightarrow\infty$) of the $ISO(3)$ Chern-Simon-Witten
theory \refs{\CSW,\TURAEV,\OS},  we  conclude
\eqn\eq{Z({L})=Z_{\rm PR}({L})=
 Z_{\rm CSW}({\cal M})\Big\vert_{(k=\infty)}~,}
where $\cal M$ is a manifold with tetrahedral decomposition $L$.

\newsec{Conclusions and Questions}

We have defined a general class of three--dimensional topological
lattice field theories, and shown that they are in one--to--one
correspondence with Hopf algebras with antipode
$S={1\over\Lambda_0}\,T$.  For the Hopf algebra $\C[G]$, we were able to
show that the corresponding TLFT is equivalent to the lattice gauge theory
with gauge group $G$ at zero coupling. Furthermore, for
$G=$SU(2), the TLFT was also equivalent to the theory of Ponzano and Regge.

A number of questions are suggested by our results. Perhaps
the most obvious is, how are our TLFTs related to other known
topological theories in three dimensions, both on the lattice and in
 continuum? For example, the Ponzano--Regge theory, and hence the
TLFT with $H=\C[$SU(2)], is known to be equivalent to the
continuum Chern--Simons--Witten theory with gauge group $ISO(3)$
\refs{\TURAEV,\OS}.  We may conjecture that
the CSW for the general inhomogeneous Lie group $IG$ is equivalent to the
TLFT with $H=\C[G]$. But that only begs the question: what TLFT is
equivalent to CSW with arbitrary gauge group $G$?

The Ponzano--Regge theory is the classical limit of the topological
lattice theory of Turaev and Viro based on the quantum group
$SU_q(2)$. Because $SU_q(2)$ is a Hopf algebra, one might expect the
Turaev--Viro theory to be equivalent to the TLFT for $H=SU_q(2)$.
However, because $S^2\ne 1$, this Hopf algebra does not obey our
constraint (except for $q=\pm 1$), so either we should look for
another Hopf algebra or for a way to relax the
constraint. The latter approach would require a generalization of our
ansatz for three--dimensional lattice field theories.

One simple generalization of our ansatz, suggested by
ref.\ \FG , would be to assign colors to faces
as well as to links.  Consider, for example, a theory
in which only the faces are colored.  Given a colored tetrahedral
lattice, we can assign a weight $C_{xyzw}$ to each tetrahedron, a
function of the colors $x,~y,~z,$ and $w$ on its four faces (ordered
with a canonical orientation). The partition function is then the
product over all tetrahedral cells of the weights, with indices
contracted using a face--gluing operator $g^{xy}$.  Topological invariance
(restricted to tetrahedral lattices) is equivalent to invariance
under the (2,3) and bubble moves, which are represented, respectively,
by the following conditions on the weights:
\eqn\aTWOTHREE{
{C_{xyz}}^u {C_{urst}} = {C_{uxr}}^v{C_{wys}}^u{C_{ztv}}^w}
\eqn\aBUB{{C_{x}}^{uvw}C_{wvuy} = g_{xy}}
Proceeding as before, we can treat the ${C_{xyz}}^w$ as structure
constants for an algebraic structure with a ``triple multiplication,''
a trilinear map $m:~A\otimes A\otimes A \to A$. Such structures are
known as triple systems and have been extensively studied\
 \TRIPLE , but to our knowledge no study has been
made of triple systems subject to the constraints \aTWOTHREE\ and \aBUB.

Another obvious direction for future work is to study TLFTs in higher
dimensions. The extension from two to three dimensions that we have
presented suggests that the most straightforward generalization to $D$
dimensions would be the following:  First we regard a lattice as consisting of
($D-1$)--simplicies glued along ($D-2$)--dimensional hinges with
($D-2$)--dimensional boundaries.  We then color these boundaries, and
assign local weights individually to both the ($D-1$)--simplicies and the
hinges, as functions of the colorings on their boundaries.  The partition
function is thus defined as the product of all the local weights
summed over all colorings.  In fact, when $D=3$, the
TLFTs we have studied are based on lattices consisting of
$2$--simplicies (triangles) glued along hinges with $1$--dimensional
boundaries.  Also, in the dual TLFTs in two dimensions
discussed in subsec.\ 2.6, we assigned weights to 1--simplices (the dual of
 gluing operators) and to zero--dimensional hinges (trivalent vertices,
{\it i.e.,} the dual of triangles)
as  functions of the colorings on their zero--dimensional boundaries.

Let us now consider a four--dimensional LFT, in which
2--simplices are colored and weights are assigned to 3--simplices, as
well as to two--dimensional hinges.  Then, just as before, the weight
of a  3D tetrahedron $C_{xyzw}$ is a function of the four colors on its
boundary faces.  Since any 3D polyhedron can be decomposed into 3D tetrahedra,
we can consistently define the weight of an arbitrary  3D polyhedron in
terms of the $C_{xyzw}$, provided that eqs.\ \aTWOTHREE\ and \aBUB\ are
satisfied.  In this case, the tetrahedral weights $C_{xyzw}$ again
define a constrained triple system.  On the other hand, the hinge
operator in this LFT  glues 3D tetrahedra  together on 2--simplices.
We denote the weight for a trivalent hinge
with  colors $x,~y$ and $z$ by $\Delta^{xyz}$,
and glue together these trivalent hinges using
$h_{xy}$ in order to build up $n$--valent hinges. As in Sec.\ 3, the
consistency of this construction will again require the
${\Delta_x}^{yz}$ to define a cosemisimple coassociative coalgebra.  We
suspect that the four--dimensional topological lattice model studied
by Ooguri\ \FOURD\ is related to a theory of this type. It will be
interesting to see whether such extensions to higher--dimensional
theories will prove to be as rich in structure as 2D and 3D TLFTs.

\bigskip
\centerline{\bf Acknowledgements}
\medskip
It is a pleasure to thank J. Distler, S. Hosono, H. Kawai,
A. LeClair, G. Moore, R. Myers,
H. Ooguri, R. Plesser, M. Rocek, S.-H. H. Tye, T. Ward and B. Zumino for
stimulating
discussions.  This work was supported in part by the National Science
Foundation.


\bigskip
\appendix{A}{Brief Review of Hopf Algebras}

In this appendix, we briefly review the basic structure of Hopf algebras \Abe.
We also prove several theorems used in sec.\ 3.
\subsec{Algebra $(H;m,u)$}
A vector space $H$ over $\C$ is called an {\bf algebra} if the two
(bi--) linear maps  {\eqnn\aMULT}  {\eqnn\aUNIT}
$$\eqalignno{
&(1)~~{\rm multiplication;}~~m:~H\otimes H \longrightarrow H&\aMULT\cr
&(2)~~{\rm unit;}~~u:~\C\longrightarrow H &\aUNIT \cr}$$
satisfy the following commutative diagrams:

(\romannumeral1) associativity:
\eqn\aASSOCDIAG{

\def\mapright#1{\smash { \mathop {\longrightarrow}\limits^{#1} }}
\def\mapdown#1{\Big\downarrow
\rlap{$\vcenter{\hbox{$\scriptstyle#1$}}$}}
\matrix{ &{H\otimes H\otimes H} &\mapright{m\otimes 1} &{H\otimes H}\cr
&\mapdown{1\otimes m}&&\mapdown{m}\cr
&H\otimes H &\mapright{m} &{H}\cr}  }

(\romannumeral2) unit:
\eqn\aUNITDIAG{

\def\mapright#1{\smash { \mathop {\longrightarrow}\limits^{#1} }}
\def\mapleft#1{\smash { \mathop {\longleftarrow}\limits^{#1} }}
\def\mapdown#1{\Big\downarrow
\rlap{$\vcenter{\hbox{$\scriptstyle#1$}}$}}
\matrix{ &{\C \otimes H} &\mapright{u\otimes 1} &{H\otimes H}
&\mapleft{1\otimes u} &{H\otimes \C}\cr
&&{\searrow^{\sim}}&\mapdown{m}&{{}^{\sim}\!\!\swarrow}&\cr
&&&H&&&\cr}  }
If we denote $m(\phi\otimes\phi')$ by $\phi\cdot\phi'$, and $u(1)$ by
$1_H$, then eqs.\ \aASSOCDIAG\ and \aUNITDIAG\ are expressed,
respectively, as
\eqn\aPHIASSOC{
(\phi\cdot\phi')\cdot\phi''=\phi\cdot(\phi'\cdot\phi'')}
\eqn\aPHIUNIT{
1_H\cdot\phi=\phi=\phi\cdot 1_H. }
For later convenience, we fix a basis of $H$ as
$H=\bigoplus_{x\in X} \C\, \phi_x$, and introduce the structure constants
${C_{xy}}^z$ and $u^x$ as
\eqn\aC{
\phi_x\cdot\phi_y~\left(\,\equiv
m(\phi_x\otimes\phi_y)\,\right)~=~{C_{xy}}^z\phi_z}
\eqn\aU{
1_H~\left(\,\equiv  u(1)\,\right)~=~u^x\phi_x~}
where repeated indices are understood to be summed over. It is easy
to show that eqs.\ \aASSOCDIAG\ and \aUNITDIAG\ (or equivalently
\aPHIASSOC\ and \aPHIUNIT ) can be rewritten as
\eqn\aCC{
{C_{xy}}^u{C_{uz}}^w = {C_{xu}}^w{C_{yz}}^u}
\eqn\aUC{u^x{C_{xy}}^z=\delta_y^z={C_{yx}}^zu^x. }

The {\bf metric} $g_{xy}$ of the algebra is defined by
\eqn\aMETRIC{
g_{xy}={C_{xu}}^v{C_{yv}}^u~,}
and the matrix $(g_{xy})$ is nondegenerate if and only if the algebra
$(H;m,u)$ is semisimple (\ie\ isomorphic to a direct sum of
matrix rings) \FHK. Furthermore, one can easily prove the following
theorem:

{\medskip
\noindent{\bf Theorem A.1.}{\it
\item{(a)}{$C_{xyz}\equiv {C_{xy}}^{z'}g_{z'z}$ is cyclically
symmetric:}
\eqn\aCYC{
C_{xyz}=C_{yzx}=C_{zxy}~.}
\item{(b)}{For $u_z\equiv g_{zz'}u^{z'}$, we have}
\eqn\aCYC{  {C_{xy}}^z u_z=g_{xy}~. }
\item{(c)}{If $H$ is semisimple, then $u^x$ is uniquely determined
 by ${C_{xy}}^z$;}
\eqn\aUGC{ u^x=g^{xx'}{C_{x'u}}^u~,}
\item{}{where $(g^{xy})$ is the inverse of $(g_{xy})$;
$g^{xy}g_{yz}=\delta^x_z$.}}
\medskip}

\subsec{Coalgebra $(H;\Delta,\e)$}
A vector space $H$ over $\C$ is called a {\bf coalgebra} if the two
(bi--) linear maps   {\eqnn\aCOMULT}  {\eqnn\aCOUNIT}
$$\eqalignno{
&(1')~~{\rm comultiplication};~~\Delta : H\to H\otimes H&\aCOMULT\cr
&(2')~~{\rm counit};~~\e : H\to \C&\aCOUNIT\cr}$$
satisfy the following commutative diagrams:

(\romannumeral1$'$) coassociativity:
\eqn\aCOADIAG{

\def\mapright#1{\smash { \mathop {\longrightarrow}\limits^{#1} }}
\def\mapleft#1{\smash { \mathop {\longleftarrow}\limits^{#1} }}
\def\mapup#1{\Big\uparrow
\rlap{$\vcenter{\hbox{$\scriptstyle#1$}}$}}
\matrix{ &{H\otimes H\otimes H} &\mapleft{\Delta\otimes 1} &{H\otimes H}\cr
&\mapup{1\otimes \Delta}&&\mapup{\Delta}\cr
&H\otimes H &\mapleft{\Delta} &{H}\cr}  }

(\romannumeral2$'$) counit:

\eqn\aCOUDIAG{

\def\mapright#1{\smash { \mathop {\longleftarrow}\limits^{#1} }}
\def\mapleft#1{\smash { \mathop {\longrightarrow}\limits^{#1} }}
\def\mapup#1{\Big\uparrow
\rlap{$\vcenter{\hbox{$\scriptstyle#1$}}$}}
\matrix{ &{\C \otimes H} &\mapright{\epsilon\otimes 1} &{H\otimes H}
&\mapleft{1\otimes \epsilon} &{H\otimes \C}\cr
&&{\nwarrow^{\sim}}&\mapup{\D}&{{}^{\sim}\!\!\nearrow}&\cr
&&&H&&&\cr}}

Fixing a basis of $H$ as $H=\bigoplus_{x\in X} \C \phi_x$, and
introducing the costructure constants ${\Delta_{x}}^{yz}$ and $\e_x$
as  {\eqnn\aDELT}  {\eqnn\aEPS}
$$\eqalignno{
\Delta(\phi_x)&={\Delta_{x}}^{yz}\phi_y\otimes \phi_z &\aDELT\cr
\e(\phi_x)&=\e_x ~, &\aEPS\cr}$$
we can rewrite eqs.\ \aCOADIAG\ and \aCOUDIAG, respectively, as
\eqn\aASS{
{\Delta_{x}}^{yu}{\Delta_{u}}^{zw}
={\Delta_{x}}^{uw}{\Delta_{u}}^{yz}}
\eqn\aDE{\
{\Delta_{x}}^{yz}\e_z=\delta_x^y=\e_z{\Delta_{x}}^{zy}.}

The {\bf cometric} $h^{xy}$ of the coalgebra is defined by
\eqn\aCOM{
h^{xy}={\Delta_{u}}^{vx}{\Delta_{v}}^{uy}.}
These $h^{xy}$, ${\Delta_{x}}^{yz}$, and $\e_x$ obey the following theorem
similar to Theorem A.1:

\medskip
\noindent{\bf Theorem A.2.}{\it
\item{(a$'$)}{$\Delta^{xyz}\equiv h^{xx'}{\Delta_{x'}}^{yz}$ is cyclically
symmetric:}
\eqn\aCYCD{
\D^{xyz}=\D^{yzx}=\D^{zxy}~.}
\item{(b$'$)}{For $\e^x\equiv h^{xx'}\e_{x'}$, we have}
\eqn\aCYCD{
\e^x {\Delta_{x}}^{yz}=h^{yz}~.}
\item{(c$'$)}{If $(h^{xy})$ is nondegenerate,
then $\e_x$ is uniquely determined by ${\Delta_{x}}^{yz}$;}
\eqn\aUGC{
\e_x={\Delta_{u}}^{ux'}h_{x'x}~,}
\item{}{where $(h_{xy})$ is the inverse of $(h^{xy})$;
$h^{xy}h_{yz}=\delta^x_z$.}}
\medskip

Note that, given an algebra $(H;m,u)$,
we can always construct a coalgebra $({\widetilde H}; \widetilde \D,\widetilde
\e)$.  In fact, if we introduce ${\widetilde H} \equiv \bigoplus_{x\in X}
\C\phi^x$, and define the comultiplication ${\widetilde \D}:~{\widetilde H}
\to  {\widetilde H} \otimes {\widetilde H}$ and the counit
${\widetilde \epsilon}:~{\widetilde H}\to \C$ as
{\eqnn\aTILDD}   \eqnn\aTILDU
$$\eqalignno{
{\widetilde \D}\,(\phi^x) &\equiv {C_{yz}}^x\phi^y\otimes\phi^z &\aTILDD\cr
{\widetilde \e}\,(\phi^x) &\equiv u^x~.&\aTILDU\cr}$$
then, as is easily shown, the  associativity and unit conditions for the
algebra $H$ imply the coassociativity and counit
conditions for the coalgebra $\widetilde H$.

Similarly, we can construct an algebra $(\widetilde H; {\widetilde m},
\widetilde u)$ from a given coalgebra $(H;\D,\,\epsilon)$ as
{\eqnn\aTILDH  \eqnn\aTILDM  \eqnn\aTILDU}
$$\eqalignno{
{\widetilde H} &=\bigoplus_{x\in X}\C\phi^x&\aTILDH\cr
{\widetilde m}(\phi^x\otimes\phi^y) &\equiv {\D_z}^{xy}\phi^z &\aTILDM\cr
{\widetilde u}(\phi^x) &\equiv \e_x \phi^x~,&\aTILDU\cr}$$

\subsec{Bialgebra $(H;m,u,\D,\e)$}

Suppose that a vector space $H$ is an algebra with respect to $(m,u)$
and also a coalgebra with respect to $(\D,\e)$. Then we can prove\ \Abe\
\medskip
\noindent{\bf Theorem A.3.}\
{\it  The following statements are equivalent:
\item{(\romannumeral1)}{$m$ and $u$ are coalgebra morphisms with respect to
$(\D,\e)$.}
\item{(\romannumeral2)}{$\D$ and $\e$ are algebra morphisms with respect to
$(m,u)$.}}
\medskip

\noindent In fact, each of these statements is equivalent to the
following set of four conditions:{\eqnn\aDM }{\eqnn\aDU}{\eqnn\aEM}{\eqnn\aEU}
$$\eqalignno{
\D\circ m &= (m\otimes m)\circ(1\otimes \tau\otimes 1)
\circ (\D\otimes \D)&\aDM\cr
\D\circ u &= u\otimes u &\aDU\cr
\e\circ m &= \e\otimes \e &\aEM\cr
\e\circ u &= 1~, &\aEU\cr}$$
where $\tau$ is the twist mapping, $\tau:~H\otimes H\owns
\phi\otimes \phi'\longmapsto \phi'\otimes\phi\in H\otimes H$.
In terms of the basis $\{ \phi_x\,;~x\in X\}$, these equations become
{\eqnn\aCD}{\eqnn\aUD}{\eqnn\aCE}{\eqnn\aEUTWO}
$$\eqalignno{
{C_{xy}}^z {\Delta_z}^{ab}
&={\Delta_x}^{pq}{\Delta_y}^{rs}{C_{pr}}^a{C_{qs}}^b&\aCD\cr
u^x{\Delta_x}^{yz}&=u^y u^z &\aUD\cr
{C_{xy}}^z \e_z&=\e_x \e_y &\aCE\cr
\e_x u^x&= 1~.&\aEUTWO\cr}$$

The quintet $(H;m,u,\D,\e)$ is called a {\bf bialgebra} if either
of the two equivalent conditions in Theorem A.3 is satisfied.
Note that the quintet $(\widetilde H;\widetilde m,\widetilde u,\widetilde \D,
\widetilde \e)$ as defined at the end of subsection A.2 is also a bialgebra if
(and only if) $(H;m,u,\D,\e)$ is a bialgebra.

\subsec{Hopf algebra $(H;m,u,\D,\e,S)$}

Let  $(H;m,u,\D,\e)$ be a bialgebra. If a linear map $S:~H\to H$
satisfies the following equation
\eqn\aANTIPODE{
m\circ (1\otimes S)\circ \D=m\circ(S\otimes 1)\circ \D=u\circ \e~,}
then the sextet  $(H;m,u,\D,\e,S)$ is said to be a {\bf Hopf algebra}
with antipode $S$. With respect to our usual basis, eq.\ \aANTIPODE\
may be expressed as
\eqn\aDSC{{\Delta_x}^{ab}{S^c}_b{C_{ac}}^y
={\Delta_x}^{ab}{S^c}_a{C_{cb}}^y=u^y\e_x }
with $S(\phi_x)={S^y}_x\phi_y$.

For the antipode $S$, we have the following two important theorems\ \Abe:

\medskip
\noindent{\bf Theorem A.4.}\
{\it The antipode $S$ is unique if it exists.
Furthermore, $S$ is an algebra antimorphism with respect to $(m,u)$:
\item{}{(\romannumeral1) $S\circ m=m\circ (S\otimes S)\circ\tau$, \ie}
\eqn\aSM{
S(\phi\cdot\phi')=S(\phi')\cdot S(\phi)~~~~{\rm for}~\phi,\phi'\in H}
\item{}{(\romannumeral2) $S\circ u=u$, \ie}
\eqn\aSU{S(1_H)=1_H,}

\noindent
and $S$ is also a coalgebra antimorphism  with respect to $(\Delta,\e)$:
{\eqnn\aDS}  {\eqnn\aES}
$$\eqalignno{
{\rm (\romannumeral3)}&~~\Delta\circ S=\tau\circ(S\otimes  S)\circ\D&\aDS\cr
{\rm (\romannumeral4)}&~~\e\circ S=\e~.&\aES\cr}  $$}
\medskip
\noindent These equations \aSM --\aES\
can be rewritten in terms of the basis $\{ \phi_x;~x\in X\}$, as
follows:  {\eqnn\aSMA}  {\eqnn\aSUA}  {\eqnn\aDSA}  {\eqnn\aESA}
$$\eqalignno{
{\rm (\romannumeral1')}
  &~~{S^z}_{z'}{C_{yx}}^{z'}={C_{x'y'}}^z{S^{x'}}_x{S^{y'}}_y&\aSMA\cr
{\rm (\romannumeral2')}
  &~~{S^x}_yu^y=u^x&\aSUA\cr
{\rm (\romannumeral3')}
  &~~{\D_{x'}}^{zy}{S^{x'}}_x={S^y}_{y'}{S^z}_{z'}{\Delta_x}^{y'z'}
  &\aDSA\cr
{\rm (\romannumeral4')}
  &~~\e_x\,{S^x}_y=\e_y~.&\aESA\cr }$$

\medskip
\noindent{\bf Theorem A.5.}\ {\it
The following conditions on the antipode $S=({S^x}_y)$ are equivalent:}
{\eqnn\aSS}  {\eqnn\aDSS}  {\eqnn\aDSCUE}
$$\eqalignno{
{\rm (\romannumeral1)}&~~S^2=1 &\aSS \cr
{\rm (\romannumeral2)}&~~{\Delta_x}^{ab}{S^c}_b{C_{ca}}^y=u^y \e_x &\aDSS \cr
{\rm (\romannumeral3)}&~~{\D_x}^{ab}{S^c}_a{C_{bc}}^y=u^y\e_x~. &\aDSCUE \cr}$$
\medskip

\noindent An immediate consequence is the following:

\medskip
\noindent{\bf Corollary A.6.}  {\it If the Hopf
algebra $(H;m,u,\D,\e,S)$ is commutative (${C_{xy}}^z=
{C_{yx}}^z$) or cocommutative (${\D_x}^{yz}={\D_x}^{zy}$), then
$S^2=1$.}
\medskip

We make one final remark concerning Hopf algebras: if $(H;m,u,\D,\e,S)$
is a Hopf algebra, then $(\widetilde H;\widetilde m,\widetilde u, \widetilde
\D, \widetilde \e, \widetilde S {\equiv\ } {^tS})$ is also a Hopf algebra,
and is called the {\bf dual Hopf algebra}.  Here, $^tS$ is the
transpose of $S$;~$^tS(\phi^x)={S^x}_y\,\phi^y$ for $\phi^x \in \widetilde H$.

\subsec{Properties of the Operator $T$ Following from the Constraint\
\hingeidGene }

Let $T :\,H\,\ni\,\phi_y~\mapsto~\phi_xT^x_{\,~y}~\,\in\,H$ be a linear
map with $T^x_{\,~y}\equiv h^{xz}g_{zy}$, and let $\Lambda_0$ be a numerical
constant defined by
\eqn\eq{
  \Lambda_0~\equiv~\epsilon_xT^x_{\,~y}u^y~=~\epsilon^xu_x\, .}
Then, we have
\medskip
\noindent {\bf Theorem A.7.}\
{\it For a bialgebra
$(H;\, m,\, u,\, \D,\, \epsilon)$, the following statements are equivalent:
\item{(a)}{ $T$ satisfies the equation (eq.\ \hingeidGene )
\eqn\aaAAA{
\Lambda\,{T^z}_y\,{\Delta_z}^{ba} = {T^a}_r\,{T^b}_s\,{\Delta_y}^{rs}}
for a numerical constant $\Lambda$.}
\item{(b)}{$T$ satisfies the equations {\eqnn\aEXTRAone} {\eqnn\aEXTRAtwo}
$$\eqalignno{
{T^x}_{y}u^y~=~\Lambda_1\,u^x~~
  &(~\Leftrightarrow~h^{xy}\,u_y~=~\Lambda_1\,u^x) &\aEXTRAone \cr
  \epsilon_x\,{T^x}_{y}~=~\Lambda_2\,\epsilon_y~~
  &(~\Leftrightarrow~\epsilon^x\, g_{xy}~=~\Lambda_2\,
                               \epsilon_y) &\aEXTRAtwo \cr}$$
for numerical constants $\Lambda_1$ and $\Lambda_2$.}
\item{(c)}{ The bialgebra is enhanced to a Hopf algebra with the antipode
\eqn\aaAAD{S={1\over{\Lambda_3}}\,T}
for a numerical constant $\Lambda_3$.}

\noindent Furthermore, if (a), (b), or (c) is satisfied,
these numerical constants $\Lambda,\,\Lambda_1,\,\Lambda_2,$ and
$\Lambda_3$ are all equal to $\Lambda_0$, and the antipode $S$ in (c)
satisfies \eqn\aaAAE{S^2=1~.}   }
\smallskip

\item{}{\noindent {\it Proof.}~~$(a)\Rightarrow\,(b):$~~
Eq.\ \aEXTRAone\  can be shown as follows: We first rewrite eq.\ \aaAAA\ as
\eqn\eq{  \Lambda\,g_{yz} \Delta^{zba}~=~T^a_{\,~r}T^b_{\,~s}\Delta_y^{~rs},}
and multiply this by $h_{ac}\,u^c\,u^y$. Then, by using eq.\ \aUD\ we have
\eqn\eq{\Lambda\,u^2\, u^b~=~u^2\, T^b_{\,~s}\, u^s~~~(u^2 \equiv u^x\,u_x).}
Since $u^2$ is not zero in general, we conclude $\Lambda\,
u^b=T^b_{\,~s}\,u^s.$
Eq.\ \aEXTRAtwo\ can be obtained just by multiplying eq.\ \aaAAA\ by
$\epsilon_a\epsilon_c\left(T^{-1}\right)^c_{\,~b}$.
Note that $\Lambda_1=\Lambda_2=\Lambda$.}
\medskip

\item{}{\noindent $(b)\Rightarrow\,(c):$~~We first note that
 $\Lambda_1=\Lambda_2=\Lambda_0$ since $\epsilon_x\,u^x=1$ by eq.\ \aEUTWO .
We then multiply eq.\ \aCD\ by $\epsilon^xu_a$, and obtain
\eqn\eq{  \Lambda_0\,\epsilon_yu^b~=~\Delta_y^{~rs}T^q_{\,~r} C_{qs}^{~~b}}
or
\eqn\aSEVEN{
  \epsilon_xu^y~=~\Delta_x^{~ab}\left({1\over\Lambda_0}\,T^c_{\,~a}\right)
C_{cb}^{~~y}.}
Similarly, by multiplying $\epsilon^yu_b$, we obtain
\eqn\eq{
  \Lambda_0\,\epsilon_xu^a~=~\Delta_x^{~pq}T^r_{\,~q} C_{pr}^{~~a}}
or
\eqn\aNINE{
  \epsilon_xu^y~=~\Delta_x^{~ab}\left({1\over\Lambda_0}\,
T^c_{\,~b}\right)C_{ac}^{~~y}.}
Eqs.\ \aSEVEN\ and \aNINE\ imply that ${1\over{\Lambda_0}}\,T$ satisfies the
 axiom of antipode (see eq.\ \aDSC).
By using the uniqueness of the antipode as stated in Theorem A.4,
we conclude that the bialgebra can be enhanced to a Hopf algebra with
antipode $S={1\over\Lambda_0}\,T$.}
\medskip

\item{}{\noindent $(c)\Rightarrow\,(a):$~~Due to Theorem A.4,
$S={1\over{\Lambda_3}}\,T$ is a coalgebra antimorphism.  This is exactly the
statement (a) with $\Lambda=\Lambda_3=\Lambda_0$.}
\medskip

\item{}{\noindent
Finally, to prove that $S^2=1$
(or equivalently $T^2 =\Lambda_0^2$), we multiply eq.\ \aaAAA\ by
$\epsilon^y h_{bc}$, and rewrite eq.\ \aEXTRAtwo\ in the form
$T^x_{\,~y}\,\epsilon^y=\Lambda_2\,\epsilon^x=\Lambda_0\,\epsilon^x$.
Then,
\eqn\eq{\eqalign{
\Lambda_0^2\, \delta_c^a~&=~h_{bc}\,T^a_{\,~r}\,T^b_{\,~s}\,h^{rs} \cr
          &=~T^a_{\,~r}\,h_{bc}\,h^{bd}\,g_{ds}\,h^{rs} \cr
          &=~T^a_{\,~r}\,T^r_{\,~c}\, . ~~~~~~~~~~{\rm (QED)} \cr}}  }

\appendix{B}{Proof of Eq.\ \eLM\ and Derivation of the Bubble and (2,3) Moves}
\global\figno=1
\def\seg#1{${\overline{#1}}$}
\def\tfig#1{{\xdef#1{Fig.\thinspace B\thinspace -\the\figno}}
Fig.\thinspace B\thinspace-\the\figno \global\advance\figno by1}

\subsec{Proof of Eq.\ \eLM}

\medskip
\noindent {\bf  Lemma B.1.}
\eqn\eq{\eqalign{
 u^x\Delta_x^{{\phantom{x}}pq}\,&=u^p\,u^q\cr
 {C_{xy}}^z\,\epsilon_z&=\epsilon_x\,\epsilon_y\cr}}
\medskip
\item{}{ {\it Proof.}~~ We prove this equation by first attaching two
triangles $C_{pab}$ and $C_{qcd}$  to $u^x\Delta_x^{{\phantom{x}}pq}$, as
shown in \tfig\Blemone a,b.  We then shrink the shaded region in \Blemone b
and obtain \Blemone c.  Topological invariance  requires
\eqn\ePF{
u^x\Delta_x^{{\phantom{x}}pq}\,C_{pab}\,C_{qcd}=g_{ab}\,g_{cd}=
u^p\,C_{pab}\,u^q\,C_{qcd}~,}
where we have used eq.\ \eunitDEFA .  Since eq.\ \ePF\ holds for arbitrary
 triangles, $C_{pab}$ and $C_{qcd}$, the first half of Lemma B.1 follows.
We prove the second part  by going to the dual lattice and using
the first part of the proof. ~~ {\rm (QED)}  }
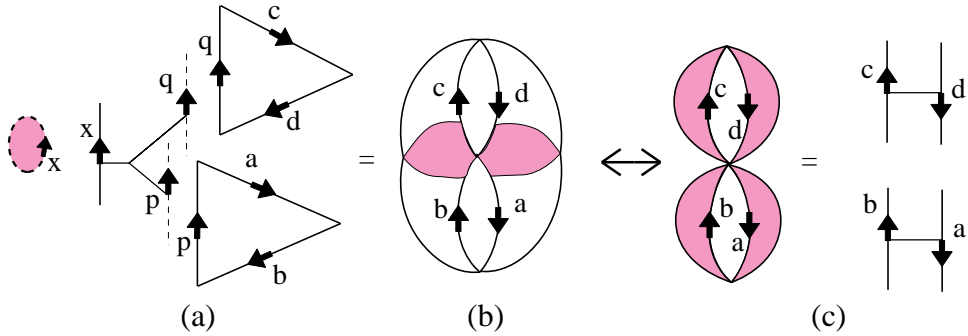
\ifig\Blemone{Proof of Lemma B.1.}{Fig-B-lem1}

\subsec{Derivation of the Bubble Move}

In this subsection, we will prove the invariance under the
bubble move, assuming only that the hinge equation  is satisfied.

\medskip
\noindent{\bf Lemma B.2.}\ {\it The configuration shown in\tfig\BA a,
consisting
of two triangles sharing three edges, is equivalent to the single triangle
appearing in \BA b.}
\ifig\BA{Lemma B.2.}{Fig-B-one}
\smallskip

\item{}{{\it Proof.}~~ The proof, shown sequentially in\tfig\BB , is
straightforward from the hinge move, \fO .  We first decompose \BB a into
a 3-hinge $\Delta_z^{{\phantom{z}}ab}$ and a configuration of
two triangles sharing two edges.
By applying the hinge move and the reduction of a   hinge loop \eDBUBBLE,
we obtain \BB e.~~(QED)}
\medskip
\ifig\BB{Proof of Lemma B.2.}{Fig-B-two}

\noindent Invariance under these moves is expressed as:
\eqn\coroI{\eqalign{
\Delta_x^{{\phantom{x}}pq}\,\Delta_y^{{\phantom{y}}rs}\,&C_{apr}\,C_{bqs}\,
\Delta_z^{{\phantom{z}}ab}=
\Lambda\,C_{xyz'}\Delta^{z'}_{{\phantom{z'}}ba}\,\Delta_z^{{\phantom{z}}ab}
\equiv \Lambda\, C_{xyz}~,\cr
&(a,b)~~~~~~~~~~~~~~~~~~~~~~~(c)~~~~~~~~~~~~~~~~~~(e)\cr}}
where the labels $(a)$-$(e)$ underlying eq.\ \coroI\ indicate the
corresponding figures in \BB .

\noindent{\bf Corollary B.3. (Bubble move):}\ {\it The configuration shown
in\tfig\BI\ consisting of two tetrahedra with three common faces is
equivalent to a single triangle.}{\phantom{\tfig\BJ}}
\ifig\BI{Bubble move.}{Fig-B-bubblemove}
\ifig\BJ{Derivation of the bubble move.}{Fig-B-PFbubble}
\smallskip

\noindent In \BI a, there are total of five triangles.
Two of them are determined by the vertices $ABC$, and share the three
edges \seg{AB}, \seg{BC} and \seg{CA}.  They  are described by
the two curved surfaces in the upper and the lower parts of the figure.
  The other three triangles are determined
by the vertices  $ABO$, $ACO$ and $BCO$.
\medskip
\item{}{ {\it Proof:}~~
We  first perform a  2D (1,3) move to replace the three internal triangles
$ABO$, $ACO$ and $BCO$ by a single triangle $ABC$ (see \BJ a).  Then
we apply {Lemma B.2} twice and obtain \BJ c.~~(QED)}
\bigskip

{\noindent{\bf Corollary B.4.}\ {\it The configuration of four triangles
linked  together as in\tfig\BC a is equivalent to the pair of triangles
shown in \BC b.}
\ifig\BC{Corollary B.4.}{Fig-B-three}  }
\smallskip

\noindent  To be more precise, we describe in words how these four triangles
are connected (see also \tfig\BD a).
The two triangles with vertices $ABC$ share  edges \seg{AC} and \seg{BC};
 the other two determined by $ABD$ join along \seg{AD} and \seg{BD}.
The two front triangles are connected  by a 3-hinge
operator along the solid edge \seg{AB}.   The two rear
triangles are connected by a 2-hinge operator along the dotted line \seg{AB}.
\medskip

\item{}{{\it Proof.}~~By applying the hinge move to
the two triangles with vertices $ABC$ in \BD b, we obtain \BD c, which contains
three different hinges, $g^{bc}$, $\Delta_u^{{\phantom{u}}da}$, and
$\Delta^{b''e{\phantom{a}}}_{{\phantom{b''e}}a}\,S^{b'}_{{\phantom{b'}}b''}\,
h_{bb'}$ (where $e$ denotes the edge variable of the triangle $C_{exy}$).
These three hinge operators can be simplified using eq.\ \eThreeC
\eqn\eSimplfy{\eqalign{
g^{bc}\,\Delta_u^{{\phantom{u}}da}\,\Delta^{b''e}_{
{\phantom{b''e}}a}\, &S^{b'}_{{\phantom{b'}}b''}\,h_{bb'}=
g^{bc}\Delta_u^{{\phantom{u}}da}\,\Delta^{b''e}_{{\phantom{b''e}}a}\,
\left( {1\over{\Lambda}}\,h^{b'w}\,g_{wb''}\,h_{bb'}\right)\cr
&= {1\over{\Lambda}}\Delta_u^{{\phantom{u}}da}\,\Delta^{ce}_{
{\phantom{ce}}a}=  {1\over{\Lambda}}\Delta_u^{{\phantom{u}}dce},\cr}}
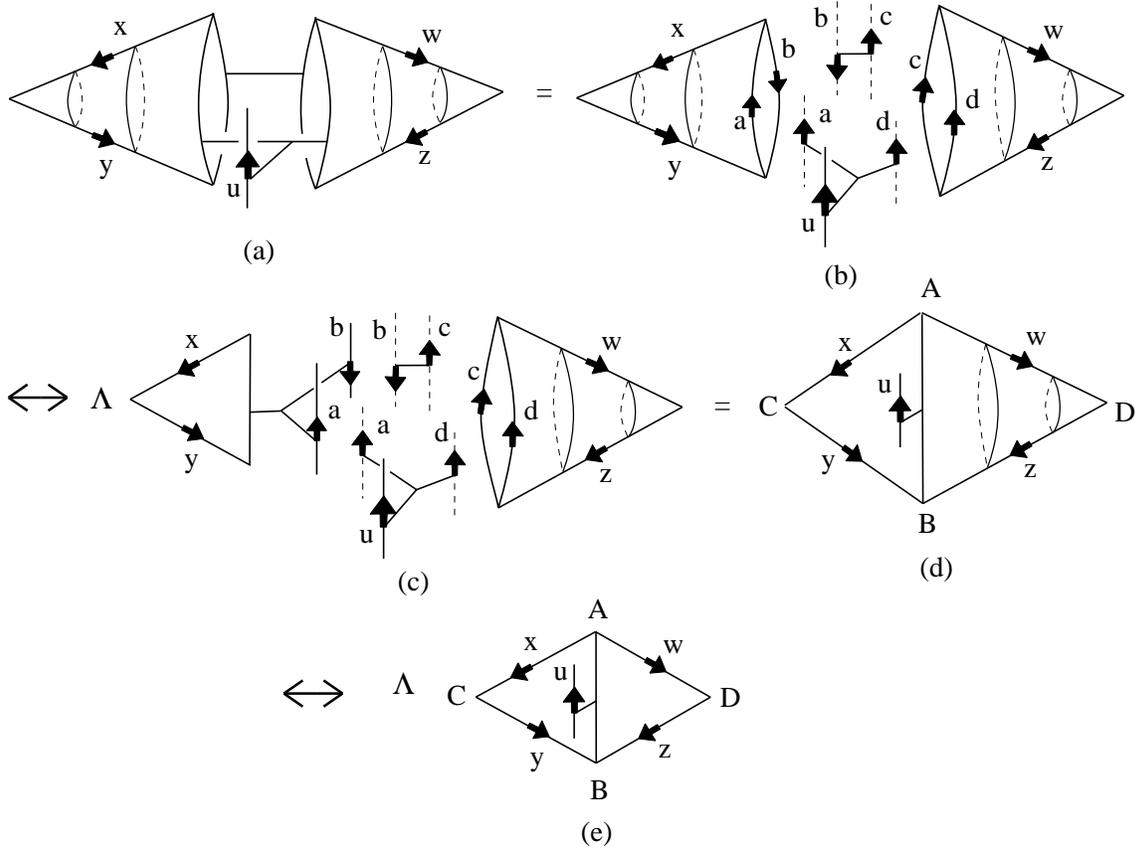
\ifig\BD{Proof of Corollary B.4.}{Fig-B-four}
which is exactly the 4-hinge operator connecting the four triangles along
the edge \seg{AB} in \BD d.  We then apply Lemma B.2 to obtain \BD e.~~(QED)}
\medskip

\subsec{Derivation of the (2,3) Move}

In this subsection, we will prove the invariance under the (2,3)
move, with the assumption that the hinge equation  is satisfied.
\medskip

\noindent{\bf Lemma B.5.} ~{\it The configuration of four triangles
in\tfig\BG a is equivalent to the three triangles in \BG b.}
\ifig\BG{Lemma B.5.}{Fig-B-six}
\smallskip

\noindent
The two triangles determined by vertices $BCD$ share the two edges
\seg{BD} and \seg{CD}.  One of these two triangles is connected to
the triangle $BCE$ by  a 2-hinge operator along the dotted line \seg{BC}.
The other triangle  is linked to
$ABC$  by a 3-hinge operator along the solid line \seg{BC}.
\medskip

\item{}{{\it Proof.}~~ We apply the hinge move to the two triangles $BCD$
in \BG a.  The resulting hinges can be  manipulated in a similar way
shown in eq.\ \eSimplfy\ to yield the hinge needed in \BG b. ~~(QED)}
\medskip

\noindent{\bf Lemma B.6.} ~{\it The three triangles join along a 3-hinge,
as shown in\tfig\BL a, is equivalent to the configurations of four triangles
shown in \BL b and \BL c.}
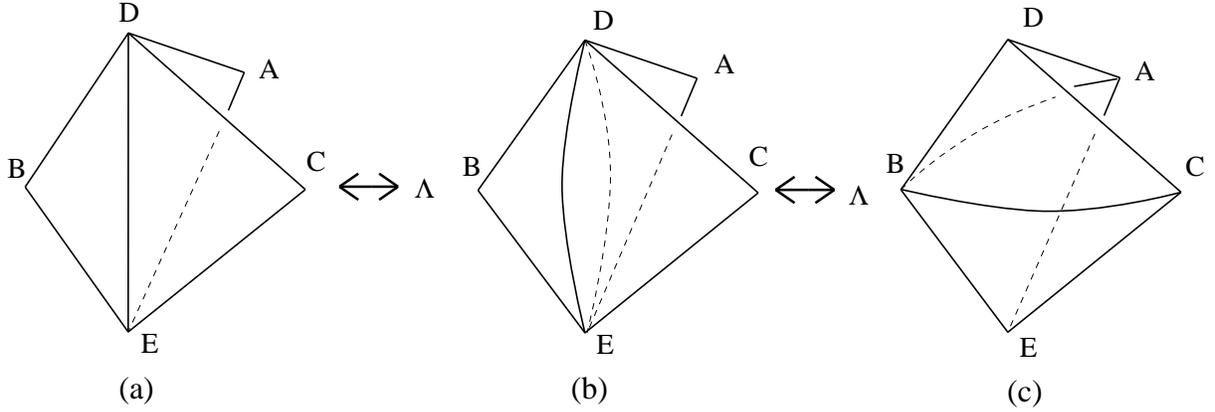
\ifig\BL{The change of three triangles joining along a 3-hinge.}{Fig-B-23two}
\medskip

\item{}{{\it Proof.}~~We first decompose the three triangles in \BL a
(or\tfig\BM a),
into the configuration in \BM b.  The hinge shown in \BM b can be written as
\eqn\eq{\eqalign{
\Delta^{vz'w'}\,S^z_{{\phantom{z}}z'}\,S^w_{{\phantom{w}}w'}
&=\Delta^{vz'w'}\,{S^{-1}}^z_{{\phantom{z}}z'}\,{S^{-1}}^w_{{\phantom{w}}w'}
=\Delta^{vz'w'}\Lambda^2\, g^{zy}\,h_{yz'}\,g^{wx}\,h_{xw'}\cr
&=\Lambda^2\,\Delta^{v}_{{\phantom{v}}yx}\,g^{zy}\,g^{wx},\cr}}
where we have used the property $S^2=1$, {\it i.e.}, $S=S^{-1}$.
$\Delta^{v}_{{\phantom{v}}yx}\,g^{zy}\,g^{wx}$ are just
 the hinges needed in \BM c.  By inflating the left triangles in
\BM c with the hinge move and connecting the resulting triangles by the
2-hinge operators, $g^{zy}$ and $g^{wx}$, we obtain  \BM e.  Then we can apply
twice the (2,2) move  to obtain \BL c.~~(QED)}
\ifig\BM{The  process from \BL a to \BL b.}{Fig-B-23three}
\medskip

\noindent{\bf Theorem B.7. ((2,3) move):}~~{\it The configuration of three
tetrahedra sharing a single edge in\tfig\BK a is equivalent to two tetrahedra
with a common face as in \BK b.}  {\phantom{\tfig\BN}}
\ifig\BK{(2,3) move.}{Fig-B-23one}
\medskip
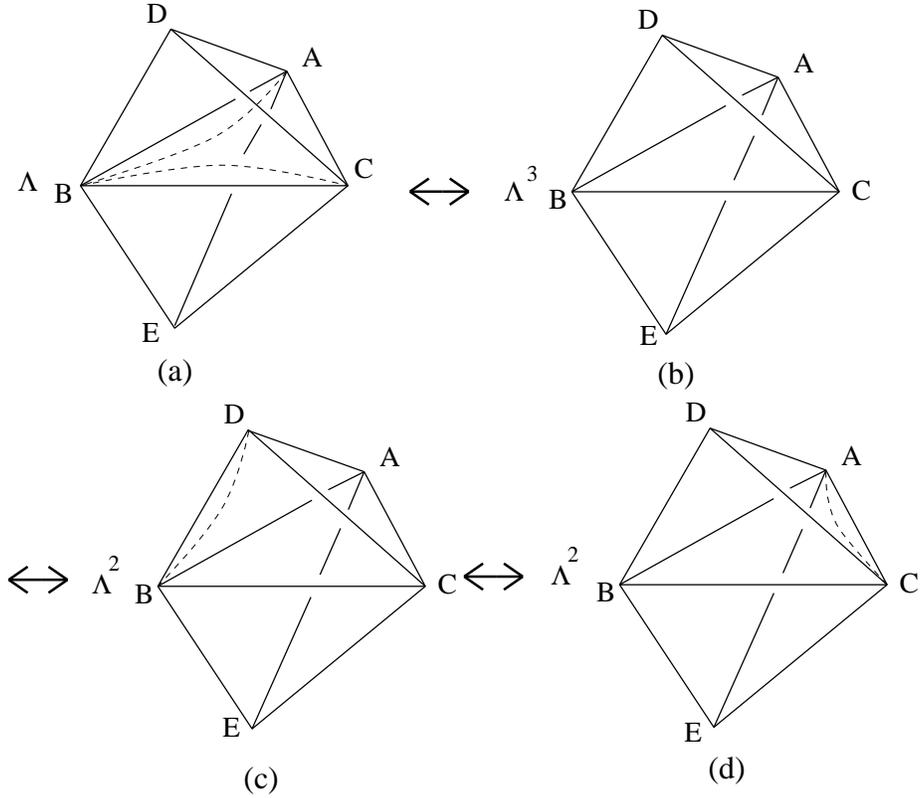
\ifig\BN{Proof of (2,3) move.}{Fig-B-23four}

\item{}{{\it Proof.}~~We first replace the three
internal triangles in \BN a by four triangles in \BL c using Lemma B.6.
The resulting four internal triangles in \BL c are connected to the original
external faces according to \BN a.  Next, we apply Lemma B.4 to deflate the
internal triangles and  obtain a
single polyhedron with six faces shown in \BN b.  Again, we use Lemma B.4
to inflate the pair of triangles $CBD$ and $BAD$, leading to the
 configuration of \BN c.  In \BN c, the two internal triangles are glued along
the dotted line \seg{BD} with a 2--hinge operator.  Then, we
apply the (2,2) move to the two internal triangles to obtain \BN d, where
the dotted line \seg{BD} has been replaced by the dotted line \seg{AC}.
Finally, we  use  Lemma B.5 to collapse the two triangles $ADC$ to a
single triangle.  This results in the configuration shown in \BK b, in which
the two tetrahedra meet on the face $ABC$.~~(QED)}
\medskip
\listrefs

\end